\documentclass[submission,Phys]{SciPost}
\usepackage{color,amsmath,amssymb,multirow,hyperref,booktabs,graphicx,mathtools,xspace,hhline} 
\newcommand{\arxiv}[1]{\href{http://arxiv.org/abs/#1}{arXiv:#1}}

\DeclareMathOperator{\sigmoid}{sigmoid}

\newcommand{\kl}{\text{KL}}

\newcommand\one{\leavevmode\hbox{\small1\normalsize\kern-.33em1}}

\newcommand{\qqquad}{\qquad \qquad}

\newcommand{\gev}{\text{GeV}}

\def\slashchar#1{\setbox0=\hbox{$#1$}           % set a box for #1
   \dimen0=\wd0                                 % and get its size
   \setbox1=\hbox{/} \dimen1=\wd1               % get size of /
   \ifdim\dimen0>\dimen1                        % #1 is bigger
      \rlap{\hbox to \dimen0{\hfil/\hfil}}      % so center / in box
      #1                                        % and print #1
   \else                                        % / is bigger
      \rlap{\hbox to \dimen1{\hfil$#1$\hfil}}   % so center #1
      /                                         % and print /
   \fi}

\newcommand{\eg}{\textsl{e.g.}\;}
\newcommand{\ie}{\textsl{i.e.}\;}

\newcommand{\be}{\begin{eqnarray*}}
\newcommand{\ee}{\end{eqnarray*}}

\newcommand{\bee}{\begin{eqnarray}}
\newcommand{\eee}{\end{eqnarray}}
\newcommand{\beeq}{\begin{equation}}
\newcommand{\eeeq}{\end{equation}}

\begin{document}

\begin{center}{\Large \textbf{
Deep-Learning Jets with Uncertainties and More
}}\end{center}

\begin{center}
Sven Bollweg\textsuperscript{1},
Manuel Hau{\ss}mann\textsuperscript{2},
Gregor Kasieczka\textsuperscript{1},\\
Michel Luchmann\textsuperscript{3}, 
Tilman Plehn\textsuperscript{3}, and
Jennifer Thompson\textsuperscript{3}
\end{center}

\begin{center}
{\bf 1} Institut f\"ur Experimentalphysik, Universit\"at Hamburg, Germany \\
{\bf 2} Heidelberg Collaboratory for Image Processing, Universit\"at Heidelberg, Germany\\
{\bf 3} Institut f\"ur Theoretische Physik, Universit\"at Heidelberg, Germany \\
plehn@uni-heidelberg.de
\end{center}

%\begin{center}
%\today
%\end{center}

% For convenience during refereeing: line numbers
%\linenumbers

\section*{Abstract}
{\bf Bayesian neural networks allow us to keep track of uncertainties,
  for example in top tagging, by learning a tagger output together
  with an error band. We illustrate the main features of Bayesian
  versions of established deep-learning taggers. We show how they
  capture statistical uncertainties from finite training samples,
  systematics related to the jet energy scale, and stability issues
  through pile-up. Altogether, Bayesian networks offer many new
  handles to understand and control deep learning at the LHC without
  introducing a visible prior effect and without compromising the
  network performance.}

\vspace{10pt}
\noindent\rule{\textwidth}{1pt}
\tableofcontents\thispagestyle{fancy}
\noindent\rule{\textwidth}{1pt}
\vspace{10pt}

\newpage
%%%%%%%%%%%%%%%%%%%%%%%%%%%%%%%%%%%%%%%%%%%%%%%%%%%%%%%%%%%%%%%%%%%%%%
\section{Introduction}
\label{sec:intro}

Modern machine learning has recently gained significant impact in many
directions of LHC physics. While boosted decision trees and relatively
simple networks have been used in particle physics for more than 30
years, new technological developments suggest using deeper networks in
a wide range of analysis tasks. The most active field in this
direction has, for a long time, been subjet physics and jet
tagging~\cite{early_stuff}. Here, multi-variate analyses of high-level
observables are currently being replaced by deep neural networks with
low-level observables. This is a natural next step given our improved
understanding of subjet physics both experimentally and theoretically,
combined with the rapid development of standard machine learning
tools. The inspiring aspect of this transition is whether deep learning
will be merely used to slightly improve existing analysis techniques,
or if we can apply it to do something new.

An early application of standard deep learning techniques to subjet
physics uses image recognition tools on so-called jet images or maps
in the rapidity vs azimuthal angle plane~\cite{jet_images}. The most
relevant information from the calorimeter then has to be combined with
tracker output, challenging at least the simplest image recognition
because of the different geometric resolution of the calorimeter and
the tracker~\cite{particlenet}.  Benchmarks processes for jet images
or alternative networks include quark-gluon
discrimination~\cite{jets_qg,lola_qg}, $W$-tagging~\cite{jets_w},
Higgs-tagging~\cite{jets_h}, or
top-tagging~\cite{deep_top1,deep_top2,lola,jets_top}. By now this
relatively straightforward classification task can be considered
solved, at least at the level of tagging
performance~\cite{ml_review,jets_comparison}. The remaining open
questions, which need to be studied before we can widely apply these kinds of
taggers to standard LHC analyses, are related to
systematics~\cite{aussies}, general uncertainties~\cite{bad}, stability, weakly
supervised learning~\cite{weak}, understanding the relevant physics
input~\cite{information}, and other LHC-specific issues which do not
automatically have a counterpart in modern machine learning.\bigskip

In particle physics applications, machine learning used for
classification will not only be judged by the best performance on a
standard data set, but by a mix of performance, properly defined
uncertainties, and stability. While, for example, including error bars
in classification tasks is not generally established~\cite{qjets}, Bayesian neural
networks (BNNs) do offer a new analysis opportunity in LHC
applications: an event-by-event estimate of the uncertainty on the
classification output. To illustrate the relevance of this
information, if a network classifies an event as 60\% signal or 40\%
background, the benefit of this piece of information rests on the
uncertainty of that per-cent value. With an error bar of $\pm 1\%$ the
60\% signal probability might well be useful for an analysis, an error
bar of $\pm 30\%$ simply means that the given event is not going to be
useful at all. Technically, Bayesian networks extract this information
by not only providing a single output variable, but a distribution of
the network output. With minimal assumptions, this distribution can
then be translated into an uncertainty band on the network output.

In this paper we will, for the first time, use Bayesian networks for
standard classification with uncertainties in the uncertainty-obsessed
field of particle physics, namely top tagging~\cite{medical}.  In
Sec.~\ref{sec:errors} we introduce Bayesian networks and discuss how
they can be applied to a simple classification task. In
Sec.~\ref{sec:useful} we relate some of their features to open
questions in particle physics applications and relate Bayesian
networks to the usual deterministic networks, where dropout and
L2-regularization are essentially Bayesian features.  For our
application we also show how the output of Bayesian networks can be
related to the frequentist approach of sampling many taggers. Finally,
in Sec.~\ref{sec:top} we test Bayesian versions of an image-based and
a 4-vector-based top tagger in a more realistic setting.  We study the
ability of the network to track uncertainties due to the limited size
of the training sample and due to systematics like the jet energy
scale. An interesting aspect is that we can separate the leading
systematic uncertainty which is correlated with a shift of the mean
network output, and a sub-leading uncorrelated systematic
uncertainty. Finally, we show how the Bayesian network offers a new
handle to test the stability of a classification network, for instance
in the presence of pile-up.

%%%%%%%%%%%%%%%%%%%%%%%%%%%%%%%%%%%%%%%%%%%%%%%%%%%%%%%%%%%%%%%%%%%%%%
\section{Machine learning with uncertainties}
\label{sec:errors}

Applying machine learning to classification tasks not only offers a way
to extremely efficiently predict properties, for example of jets, it
also allows us to define jet-by-jet uncertainty estimates on the
tagging output. Possible sources of uncertainty in a particle physics
framework include
\begin{itemize}
\item finite, but perfectly labeled training samples. In LHC analyses
  this corresponds to a statistical uncertainty in the classification
  output, for instance due to finite MC statistics;
\item inconsistencies in the training data and their labels. In LHC
  analyses those correspond to systematic uncertainties on the
  classification or tagging output;
\item differences between the training and test samples. In LHC
  analyses they arise from Monte Carlo simulation or control regions
  in data to the signal region and would again be treated as
  systematics at the classification output level.
\end{itemize}
The deep learning literature~\cite{deep_errors} defines two kinds of
uncertainty: (i) epistemic or model uncertainty describes the lack of
statistics and can therefore be expected to decrease with more data;
(ii) aleatoric errors from noise in the data, which cannot be reduced
by using more data. It makes sense to separate homoscedastic
(universal) and heteroscendastic (input-dependent) noise.  If we look
at the scaling with increased data sets, this distinction corresponds
exactly to statistical and systematic uncertainties in the LHC conventions.

It is crucial to notice that all three sources of uncertainties listed
above are induced by either not perfect training data or by a not
perfect match between training data and testing data. They are not
uncertainties on the form of the actual network, which we simply
consider a mathematical relation between network input and network
output. Nevertheless, they all describe statistical or systematic
uncertainties on the tagging output which have to be considered in the
actual analysis.

%%%%%%%%%%%%%%%%%%%%%%%%%%%%%%%%%%%%%%%%%%%%%%%%%%%%%%%%%%%%%%%%%%%%%%
\subsection{Bayesian neural networks}
\label{sec:bnn}

Like all classifying neural networks, BNNs~\cite{bnn_early,bnn_tev,bnn_nu}
relate training data $D$ to a known output or classifier $C$ through a
set of network parameters $\omega$. Bayes' theorem then defines the
(posterior) probability distribution for the parameters $p(\omega | \{
D,C \} )$ from the general relation
\begin{align}
p(\omega | \{ D,C \} )\; p( \{D,C \} ) = p( \{D,C \} | \omega) \; p(\omega) \; ,
\label{eq:bayes1}
\end{align}
In this form $\{ D,C\}$ describes the combination of training inputs
and network outputs for fully supervised learning. If we consider the
training data $D$ as given, we can omit it in the shorter form
\begin{align}
%p(\omega | D,C )\; p( C | D ) = p( C | D,\omega) \; p(\omega) 
%\quad \Leftrightarrow \quad 
p(\omega | C ) = \frac{p( C | \omega) \; p(\omega)}{p(C)} \; .
\label{eq:bayes2}
\end{align}
We can think of the prior $p(\omega)$ as the distribution of the model
parameters before training on the data set $D$ and are free to choose
it, for example, to be a Gaussian. The model evidence $p(C)$ serves
as a normalization constant for the (posterior) probability
distribution $p(\omega | C )$. The probability $p(\omega | C )$ allows
us to predict the network output $c^*$ for a new test data point,
\begin{align}
p(c^* | C)
= \int d \omega \; p(c^* | \omega, C) \; p(\omega | C )
\label{eq:folding}
\end{align}
This line of argumentation immediately leads us to the main question behind
this paper: can we define and determine a network output which is not
just one number, like a signal probability, but a signal probability
distribution in $\omega$ on a jet-by-jet level?\bigskip

The technical problem with Eq.\eqref{eq:folding} is that we usually do
not know the closed form of $p(\omega | C)$, even if it is implicitly
encoded in our neural network. On the other hand, we can approximate
it in the sense of a distribution and combine with a test function
$p(c^* | \omega)$~\cite{blei},
\begin{align}
 \int d \omega \; p(c^* | \omega) \; p(\omega | C )
\approx \int d \omega \; p(c^* | \omega) \; q(\omega) \; .
\end{align}
The agreement between $p(\omega | C )$ and such an approximation
$q(\omega)$ is given by the Kullback-Leibler divergence,
\begin{align}
\kl [q(\omega),p(\omega|C)] 
&= \int d\omega \; q(\omega) \; \log \frac{q(\omega)}{p(\omega|C)} \; .
\label{eq:kl_def}
\end{align}
It vanishes if the two functions are identical almost everywhere and
is positive otherwise.  We can use Bayes' theorem to re-write it as
\begin{align}
\kl [q(\omega),p(\omega|C)] 
&= \int d\omega \; q(\omega) \; \log \frac{q(\omega) p(C)}{p(C|\omega) p(\omega)} \notag \\
&= \kl[q(\omega),p(\omega)] 
 + \log p(C) \int d\omega \; q(\omega) 
 - \int d\omega \; q(\omega) \; \log p(C|\omega) \; .
\label{eq:kl_bayes}
\end{align}
The second term only includes the normalization of $q(\omega)$ and is
of no particular interest given that we have normalized $q(\omega)$ as
a probability distribution. The third term is the usual expected
likelihood, which we can use to work with in a frequentist sense, if
we want to avoid the Bayesian prior altogether. In our framework we
minimize the KL-divergence to construct a $q(\omega)$ approximating
$p(\omega | C )$ in Eq.\eqref{eq:folding}, so our loss function
becomes
\begin{align}
L 
& = \kl[q(\omega),p(\omega)]  
    - \int d \omega \; q(\omega) \; \log p(C|\omega) \; .
\label{eq:maximize}
\end{align}
In pushing $L$ to its well-defined lower limit, the first term
requires $q(\omega)$ to be close to an assumed, for instance Gaussian,
prior $p(\omega)$. If we assume that $q(\omega)$ and $p(\omega)$ are
both Gaussians described by their respective $\mu$ and $\sigma$ and we
only consider a single weight, we can simplify the KL-divergence to
\begin{align}
\kl[q(\omega),p(\omega)]  
=   \log \frac{\sigma_p}{\sigma_q} 
   + \frac{\sigma_q^2+ (\mu_q - \mu_p)^2}{2\sigma_p^2} 
   - \frac{1}{2}\; .
\end{align}
The last term in Eq.\eqref{eq:maximize} needs to be minimized once we
know the likelihood $p(C|\omega)$ for a given $C$ and evaluated as a
function of $\omega$.  Technically, this requires a variation of the
parameters which give the functional form of the Gaussian
$q_{\mu,\sigma}$.  That means that we compute the derivative of $L$
with respect to $\mu$ and $\sigma$, leading for example to the
condition
\begin{align}
\frac{\partial}{\partial \mu} 
  \int d \omega \; q_{\mu, \sigma}(\omega) \; \log p(C|\omega) = 0 \; .
\label{eq:minimize}
\end{align}
Once we determine the approximate probability distribution $q_{\mu,
  \sigma} (\omega)$ from Eq.\eqref{eq:maximize}, we can use it to
solve Eq.\eqref{eq:folding} by Monte Carlo integration and find the
predictive mean for the test sample,
\begin{align}
p(c^* | C)
%= &  \int d \omega \; p(c^* | \omega) \; p(\omega | C )\\
\approx  \int d \omega \; p(c^* | \omega) \; q_{\mu,\sigma}(\omega) 
\approx  \frac{1}{N} \sum_j^N \; p(c^* | \omega_j \big(\mu,\sigma) \big) 
\equiv \mu_\text{pred} \; .
\label{eq:pred_mean}
\end{align}
To compute this mean we use $N$ sets of weights $\{ \omega \}$,
effectively corresponding to $N$ networks with different weights.
Assuming a Gaussian probability distribution we also need the spread
of the $N$ sets of weights, or the predictive standard deviation
\begin{align}
\sigma_\text{pred}^2 
= \frac{1}{N} \sum_{j}^{N} \big[ p(c^*|\omega_j(\mu,\sigma)) - \mu_\text{pred} \big]^2 \; .
\label{eq:pred_std}
\end{align}
This way the BNN returns not only a central value $\mu_\text{pred}$
for the classifying outcome, but also an jet-by-jet uncertainty
estimate for this classification outcome $\sigma_\text{pred}$.

%%%%%%%%%%%%%%%%%%%%%%%%%%%%%%%%%%%%%%%%%%%%%%%%%%%%%%%%%%%%%%%%%%%%%%
\subsection{Probabilities}
\label{sec:classification}

%------------------------------------------------
\begin{figure}[t]
\centering
\includegraphics[width=0.49\textwidth]{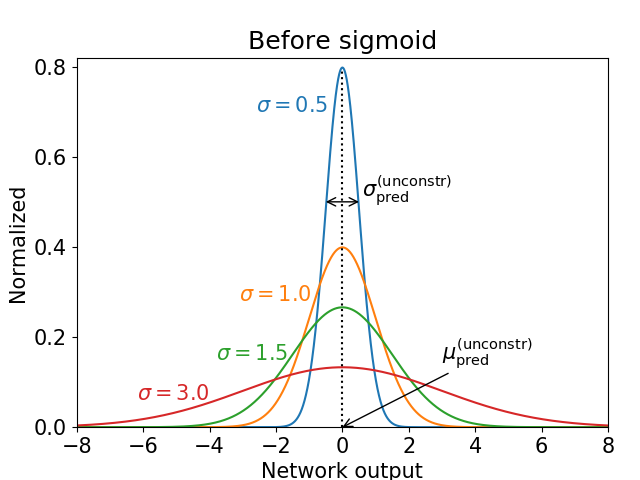}
\includegraphics[width=0.49\textwidth]{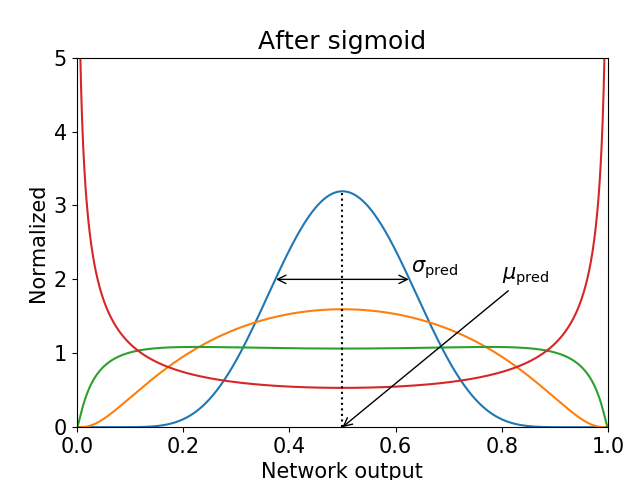} \\
\includegraphics[width=0.49\textwidth]{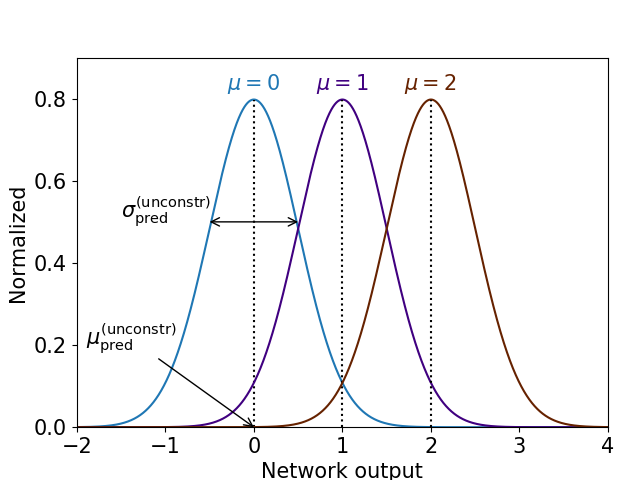}
\includegraphics[width=0.49\textwidth]{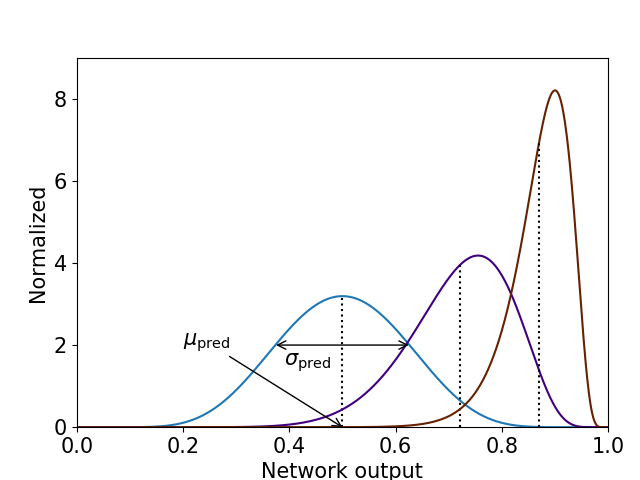} 
\caption{Effect of the sigmoid transformation on normal distributions
  with the same mean but different widths (upper) and on the same width
  but different means (lower).}
\label{fig:sigmoid1}
\end{figure}
%------------------------------------------------

We can numerically test this behavior with a toy BNN, analyzing jet
images with $40\times 40$ pixels. It does not include a convolutional
layer and only consists of two fully connected hidden layers with a ReLu
activation function, each with 100 units, and an output layer with one unit and a sigmoid activation function. It delivers a scalar output.
For the BNN version of this toy network we use the TensorFlow
Probability library~\cite{tensorflow_probability} and its DenseFlipout
Layer~\cite{flipout}. We have convinced ourselves that sampling 100
times from the weight distributions gives us stable results.  Unless
specified otherwise, we train our toy model on 100k top and 100k QCD
jets. The BNN property means that we are not only interested in the
values of the network output, but in the output distributions, as we
will discuss in Sec.~\ref{sec:inside}. Our toy BNN is trained to
distinguish a public set of 600k top jet and QCD jet images
each~\cite{lola}, which were generated with
\textsc{Pythia8}~\cite{pythia} for an LHC energy of 14~TeV and without
pile-up or multiple interactions. As a simplified detector simulation
we use \textsc{Delphes}~\cite{delphes} with the default ATLAS detector
card. The fat jet is defined through the anti-$k_T$
algorithm~\cite{anti_kt} in \textsc{FastJet}~\cite{fastjet} with
$R=0.8$, fulfills
\begin{align}
p_{T,j} = 550~...~650~\gev
\qquad \text{and} \qquad 
|\eta_j|<2 \; .
\end{align}
The top jets are truth-matched to a $b$-quark and two light quarks
within $\Delta R=0.8$.  The images include the improved pre-processing
taken from Ref.~\cite{deep_top2}.

%------------------------------------------------
\begin{figure}[t]
\includegraphics[width=0.325\textwidth]{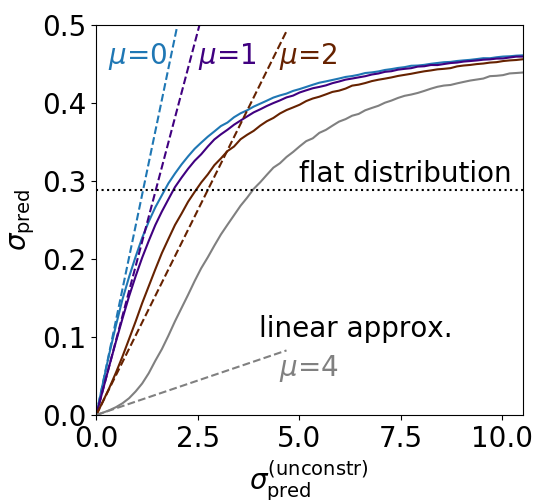}
\includegraphics[width=0.325\textwidth]{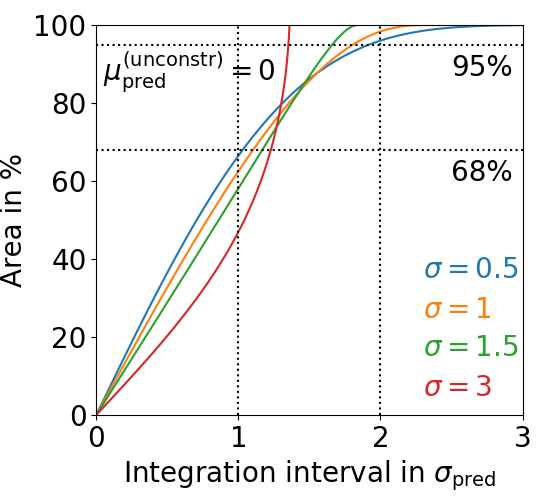}
\includegraphics[width=0.325\textwidth]{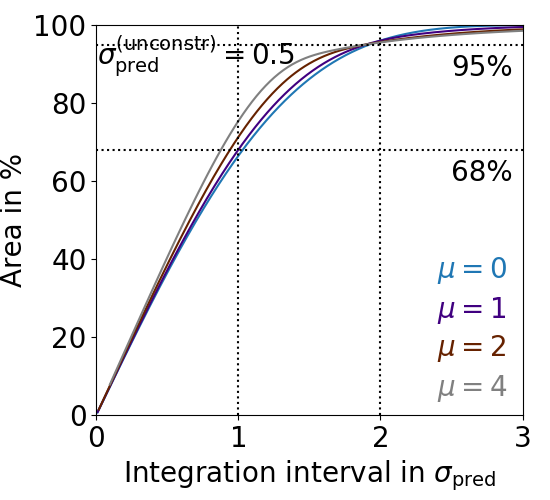}
\caption{Left: predictive standard deviation of the after sigmoid
  transformation, the linear approximation is defined in
  Eq.\eqref{eq:sigmoid_width_approx}. Center and right: area under
  the probability distribution in the interval $[\mu-x, \mu +x]$ for
  different widths and means.}
\label{fig:sigmoid2}
\end{figure}
%------------------------------------------------

Whenever we use a neural network on classification tasks with
probabilities as outputs we need to map the unbounded space of network
outputs to the closed interval $[ 0,1 ]$, for instance through a sigmoid
function
\begin{align}
\sigmoid (x) =\dfrac{ e^x}{1+e^x} = \frac{1}{1 + e^{-x}} 
\qquad \Leftrightarrow \qquad
\sigmoid^{-1} (x) = \log \frac{x}{1-x} \; .
\label{eq:sigmoid_general}
\end{align}
Because the sigmoid is non-linear, it would not maintain a Gaussian
shape of an output distribution.  In Fig.~\ref{fig:sigmoid1} we show
how increasingly larger widths and shifts in the mean lead to
non-Gaussian features in the probability distributions on the interval
$[0,1]$ for this toy network. To compute the mean including a sigmoid
transformation of a Gaussian probability distribution
$G_{\mu,\sigma}(\omega)$ we solve the integral
\begin{align}
\mu_\text{pred} 
&= \int_{- \infty}^\infty d \omega \; \sigmoid(\omega) \; G_{\mu,\sigma}(\omega) \notag \\
    &= \int_0^1 dx \; \frac{x}{x(1-x)} \; G_{\mu,\sigma}\left( \log \frac{x}{1-x} \right)  \; .
\end{align}
This probability distribution is known as the logit-normal
distribution. For small widths of the unconstrained distributions
before sigmoid, $\sigma_\text{pred}^\text{(unconstr)}$, we can approximately
calculate the predictive standard deviation after the sigmoid
transformation as
\begin{align}
\sigma_\text{pred}
\approx \mu_\text{pred} \left( 1- \mu_\text{pred} \right) \; 
        \sigma_\text{pred}^{(\text{unconstr})} 
\qqquad \text{with} \quad \mu_\text{pred} \in [0,1] \; .
\label{eq:sigmoid_width_approx}
\end{align}
In the left panel of Fig.~\ref{fig:sigmoid2} we show the correlation
between the standard deviations before and after sigmoid for different
means, including the linearized approximation shown above. For large
values of the pre-sigmoid standard deviation the behavior of the
probabilistic outcome deviates significantly from the linearized
form. This is simply an effect of the interval $[0,1]$, where we need
to keep in mind that the standard deviation of a flat distribution is
$1/(2\sqrt{3})$ and the largest possible standard deviation comes from
a bi-polar distribution with $\sigma = 1/2$. This is exactly the
plateau value we observe for $\sigma_\text{pred}$ for
large pre-sigmoid values of $\sigma_\text{pred}^{(\text{unconstr})}$.

One of the questions that arises once we attempt a frequentist
interpretation of the probability distributions from a BNN is the
relation between the deviation from the mean in terms of standard
deviations and in terms of the area under the probability
distribution. We show this correlation in the center and right panels
of Fig.~\ref{fig:sigmoid2}. The curves indicate that below the plateau
value of $\sigma_\text{pred} = 1/2$ the one-sigma and two-sigma limits
scale reasonably well with 68\% and 95\% of the full integral.

Altogether, we still need to remember that the mapping from the
network output over the space of real numbers to the interval $[0,1]$
leads to non-Gaussian probability distributions. This is similar to
the case of small event counts, where a Poisson distribution deviates
from the symmetric Gaussian because it avoids negative event
counts. Even with such non-Gaussian output the distributions before
the sigmoid transformation are Gaussian, which means that they are
fully described by the predictive mean and standard deviation. The
network output after sigmoid is also fully described by two
parameters, in our case the predictive mean and the predictive
standard deviation.

%%%%%%%%%%%%%%%%%%%%%%%%%%%%%%%%%%%%%%%%%%%%%%%%%%%%%%%%%%%%%%%%%%%%%%
\subsection{Prior (in)dependence}
\label{sec:prior}

As for any Bayesian setup, it is crucial that we confirm that the
output of our BNN top tagger is not dominated by the prior.  For the
network training, the prior enters as the function $p(\omega)$ in the
loss function.  The trained Gaussian $q_{\mu,\sigma}$ is then
constructed to fulfill the different conditions entering the loss
function, Eq.\eqref{eq:maximize}.  The default prior for our toy model
is is a normal distribution with $\mu_\text{prior}=0$ and
$\sigma_\text{prior}=1$. To test prior independence we stick to normal
distributions with $\mu_\text{prior}=0$, but vary the width of the
prior distribution over an extremely wide range,
\begin{align}
\sigma_\text{prior} = 10^{-2}~...~1000 \; .
\end{align}
The setup of our classification network is described in
Sec.~\ref{sec:classification}.  The performance of the different
networks is measured in term of the area under the ROC curve (AUC). We
find

\begin{center}
\begin{tabular}{c|cccccc}
\toprule
$\sigma_\text{prior}$ & $10^{-2}$ & $10^{-1}$ & 1 & 10 & 100 & 1000 \\ 
\midrule
AUC & 0.5 & 0.9561 & 0.9658 & 0.9668 & 0.9669 & 0.9670 \\ 
error & ---  & $\pm 0.0002$& $\pm 0.0002$& $\pm 0.0002$& $\pm 0.0002$& $\pm 0.0002$ \\
\bottomrule
\end{tabular} 
\end{center}

\noindent Indeed, a too narrow prior distribution does not allow the
network to efficiently grasp the features of the training data. As a
result, the network with $\sigma_\text{prior} = 10^{-2}$ does not
perform at all. The situation improves towards $\sigma_\text{prior} =
1$ and reaches a plateau above this value.  To judge the significance
of the change we estimate the spread of the AUC values by training our
default models five times and give the standard deviation of the
corresponding AUC values. In terms of the AUC value we find a very
slight systematic increase for larger prior widths, but at the price
of vastly increased training time. For our toy model we have checked
that also a jet-by-jet comparison of the different priors confirms the
approximate prior independence of the predictive mean and the
predictive standard deviation given by our default setup.

%%%%%%%%%%%%%%%%%%%%%%%%%%%%%%%%%%%%%%%%%%%%%%%%%%%%%%%%%%%%%%%%%%%%%%
\section{Useful features}
\label{sec:useful}

From the construction of BNNs it is obvious that modified versions of
top taggers will have some unique and attractive features for
applications in LHC physics. Still based on our toy tagger we will
illustrate three such features: the proper treatment of statistical
uncertainties from finite training samples, an established way to
calibrate the network output relative to the accuracy on a test
sample, and a proper framework of some ad-hoc features in
deterministic networks.

%%%%%%%%%%%%%%%%%%%%%%%%%%%%%%%%%%%%%%%%%%%%%%%%%%%%%%%%%%%%%%%%%%%%%%
\subsection{Statistical uncertainty from training}
\label{sec:statistical}

%------------------------------------------------
\begin{figure}[t]
\centering
\includegraphics[width=0.49\textwidth]{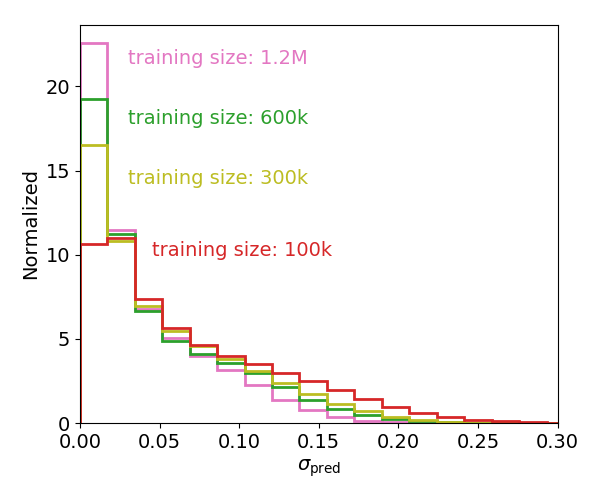}
\caption{Normalized predictive standard deviation of signal jets
  for different sizes of the training sample.}
\label{fig:sd_sig_bg} 
\end{figure}
%------------------------------------------------

To see how the the BNN setup with the predictive standard deviation
works we can look at a simple source of statistical uncertainties,
namely a limited number of training jets. Again, we use the fast and
simple toy network to describe the main features. In
Fig.~\ref{fig:sd_sig_bg} we show the normalized distribution of the
predictive standard deviations. Running the network on the test sample
of 200k top jets we histogram all predictive standard deviations for
the full range of predictive means. Indeed, the distribution becomes
more and more peaked for an increased training sample.

%------------------------------------------------
\begin{figure}[t]
\includegraphics[width=0.49\textwidth]{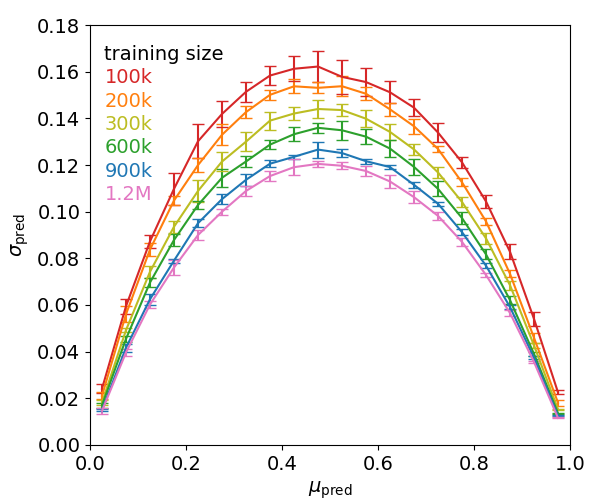}
\includegraphics[width=0.49\textwidth]{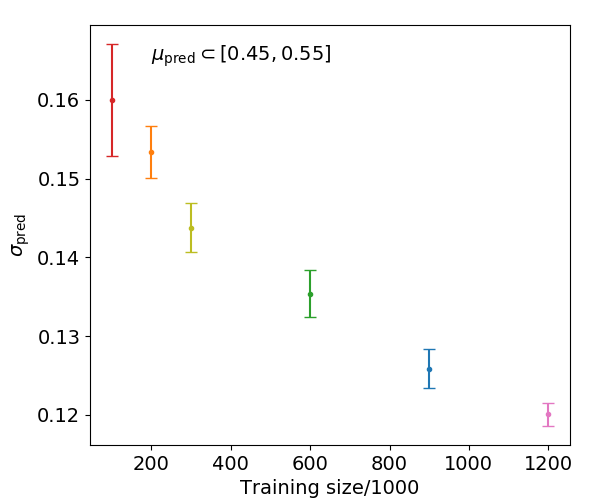} \\
\includegraphics[width=0.24\textwidth]{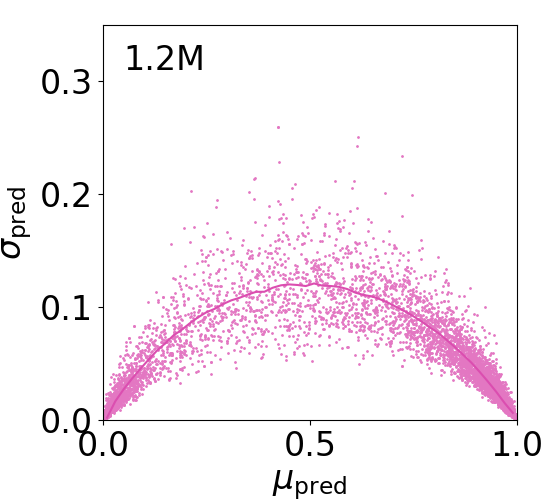}
\includegraphics[width=0.24\textwidth]{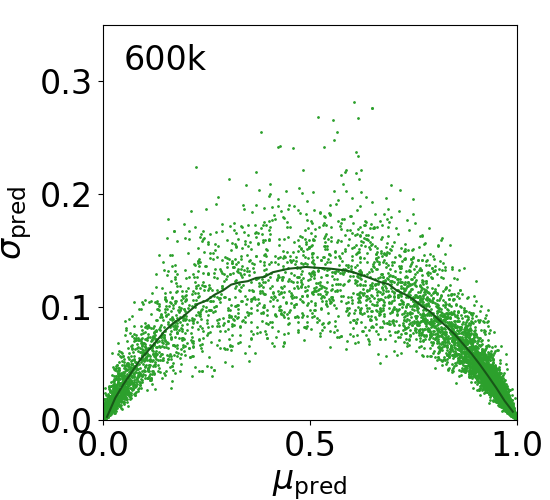}
\includegraphics[width=0.24\textwidth]{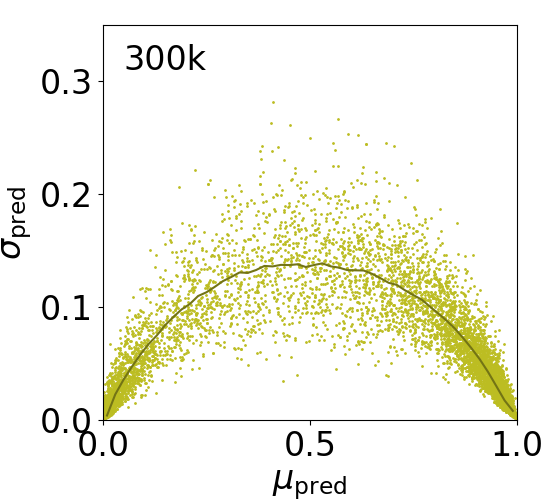}
\includegraphics[width=0.24\textwidth]{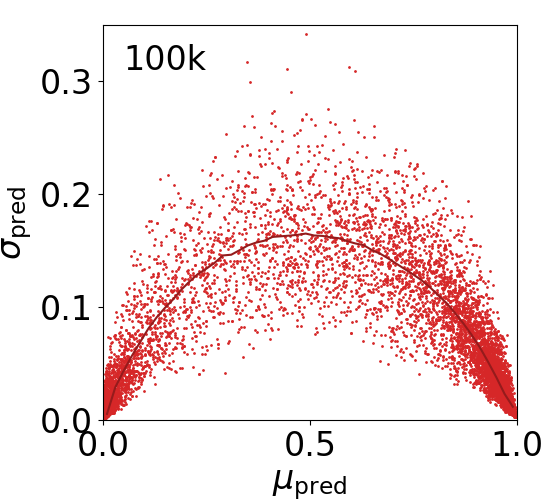}
\caption{Correlation between predictive mean and standard
  deviation. The error bars in the upper left panel correspond to five
  independent trainings and indicate the uncertainty on the
  uncertainty given by the BNN. The right panel shows the predictive
  standard deviation for $\mu_\text{pred} = 0.45~...~0.55$ as a
  function of the size of the training sample with the same error bars
  from different trainings. The lower panels instead show the
  statistical spread for 10k jets, signal and background combined.}
\label{fig:mu_sd}
\end{figure}
%------------------------------------------------

In the upper left panel of Fig.~\ref{fig:mu_sd} we show the
correlation between the predictive mean and the predictive standard
deviation as the two outputs of the BNN. To construct a single
correlation curve we evaluate the network on 10k jets, half of them
signal and half of them background. We show the mean values of the 10k
jets in slices of $\mu_\text{pred}$, after confirming that their
distributions have the expected Gaussian-like shape.  The leading
feature is an inverse parabola shape, which is induced by the sigmoid
transform, Eq.\eqref{eq:sigmoid_width_approx}. It simply reflects the
fact that a network output in the interval $[0,1]$ forces the error
bars close to the ends to be comparably small. Another, physical
source of the same effect is that probability outputs around 0.1 or
0.9 correspond to clear cases of signal and background jets, where we
can expect the predictive standard deviation, or the error on the
predictive mean, to be small.  The error bars shown in the upper
panels of Fig.~\ref{fig:mu_sd} show the uncertainty on the predictive
uncertainty. We derive them from five independent trainings and
testings, including statistically independent samples.

In the upper right panel we illustrate the improvement of the network output
with an increasing amount of training data by showing the predictive
standard deviation for $\mu_\text{pred} = 0.45~...~0.55$ as a function
of the size of the training sample.  The estimated uncertainty on the
tagger output decreases monotonically from 16\% to 12\% when we
increase the training sample from 100k to 1.2M jets. This improvement
is significant compared to the error bars, which correspond to different
training and testing samples.

Finally, the spread of these 10k signal and background jets is
illustrated in the four lower panels, with a matching color code.  We
immediately see that the spread is strongly reduced for larger
training samples.

%%%%%%%%%%%%%%%%%%%%%%%%%%%%%%%%%%%%%%%%%%%%%%%%%%%%%%%%%%%%%%%%%%%%%%
\subsection{In-situ calibration of weight distribution}
\label{sec:inside}

Before we attempt to compare the output of the BNN to a frequentist
distribution of many deterministic neural networks we can apply a
cross check within the Bayesian framework itself and construct a
hybrid version of the BNN. This will lead us to another attractive
feature of such networks, their explicit calibration based on training
data.

The standard BNN constructs its output distribution by sampling the
individual weights of each layer. They are initialized as a random set
of Gaussians with different means and standard deviations, which are then
learned during training. This means that the BNN not only learns a set
of weights, but a distribution of weights from the first layer to the
network output.

To generate a distribution we can also train the BNN a large number of
times and only use the maximum values or means of the (Gaussian)
weight distributions.  This should encode the same information as the
BNN, just much less efficiently.  One of the problems with this
so-called maximum-a-posteriori (MAP) approach is that it does not
include a Bayesian integral over weight space and instead works like a
frequentist profile likelihood. This profiling does not maintain the
normalization of the probability distribution and therefore makes it
impossible to naively compare for example ranges of the network output
of the kind $60~...~70\%$ vs $80~...~90\%$ top jet probability. To
re-gain this interpretation and to be able to compare the full BNN with
this MAP approximation we can re-calibrate the
tagger~\cite{recalibrate}. This re-calibration ensures that the
probabilistic output of the network and the measured accuracy on a
test sample are aligned, \ie an $80\%$ top jet probability network
correctly identifies 80\% of top jets in a given sample. Obviously, it
is a tool which can be applied much more generally than for our toy
comparison.

%------------------------------------------------
\begin{figure}[t]
\centering
\includegraphics[width=0.5\textwidth]{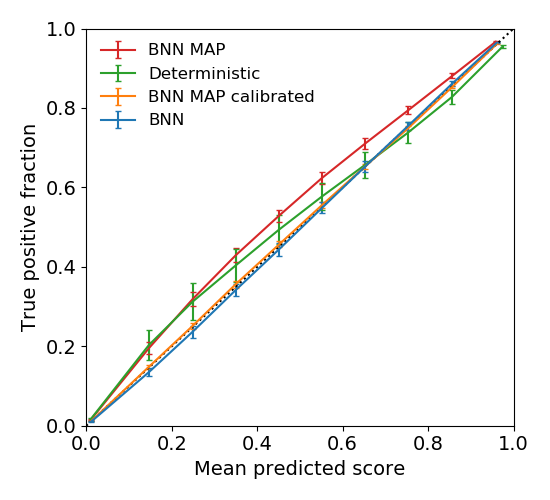}
\caption{Reliability diagrams of different BNN approaches, as well as
  the set of deterministic taggers discussed in Sec.~\ref{sec:link}.
  The bin-wise true positive rate is plotted against the mean
  prediction of the jets in one bin. The error bars indicate the
  spread for 50 independent training samples.}
\label{fig:calibration}
\end{figure}
%------------------------------------------------

The effect of such a re-calibration can be shown in reliability
diagrams like in Fig.~\ref{fig:calibration}. They are constructed by
discretizing the tagger output, \ie the signal or background
probability, into several bins of equal length and correlating this
number with the true positive rate in each bin. In a well-calibrated
model the resulting curve would sit on the diagonal, modulo
uncertainties. Using 400k top and QCD jets integrated over ten bins we
see that unlike the consistent BNN output, the MAP output is poorly
calibrated. The error bars indicate the spread for 50 independent
training samples.  We find that it can be significantly improved
through Platt scaling~\cite{platt_scaling}: here we first transform
the before-sigmoid output linearly, $x^{\prime} = a\,x + b$. The free
parameters $a$ and $b$ can then be determined by minimizing the usual
cross entropy for fixed weights on a validation set. Indeed,
the re-calibrated MAP network is well calibrated and can be used to
extract a sensible tagging output. Obviously, this kind of
re-calibration between the network output and the measured purity is
not only useful to study the behavior of our toy BNN, but it can be
used to calibrate any classification tool, for instance at the
LHC.

\begin{figure}[t]
\centering
\includegraphics[width=0.49\textwidth]{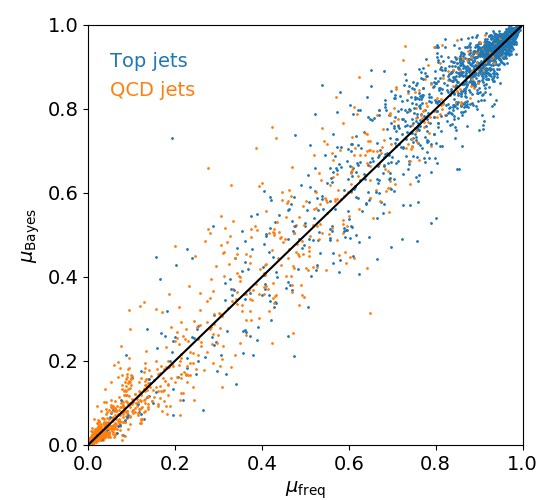}
\includegraphics[width=0.49\textwidth]{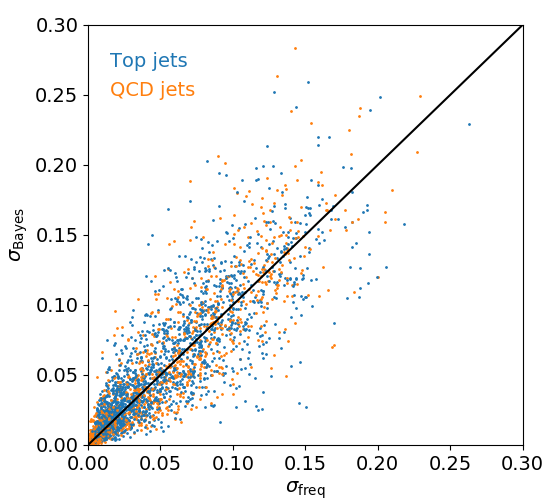}
\caption{Correlation of the mean (left) and standard deviation
  (right) between the BNN's $\mu_\text{pred}$ and $\sigma_\text{pred}$
  and the histogrammed weights fixed to their respective means. Each
  point corresponds to one of 2000 signal and 2000 background jets, as
  indicated by the colors.}
\label{fig:map} 
\end{figure}

In Fig.~\ref{fig:map} we show the correlation between the predictive
mean and predictive significance from the BNN output and using the MAP
results. Each point in the scatter plot corresponds to one of 4000
jets. Indeed, the predictive BNN output scatter the same way as the
weights after re-calibration. This implies that, in spite of the
Bayesian setup, we can think of the predictive standard deviation as
representing the distributions of the network weights in a frequentist
sense.

%------------------------------------------------
%------------------------------------------------

%%%%%%%%%%%%%%%%%%%%%%%%%%%%%%%%%%%%%%%%%%%%%%%%%%%%%%%%%%%%%%%%%%%%%%
\subsection{Relation to deterministic networks}
\label{sec:link}

%------------------------------------------------
\begin{figure}[t]
\centering
\includegraphics[width=1.\textwidth]{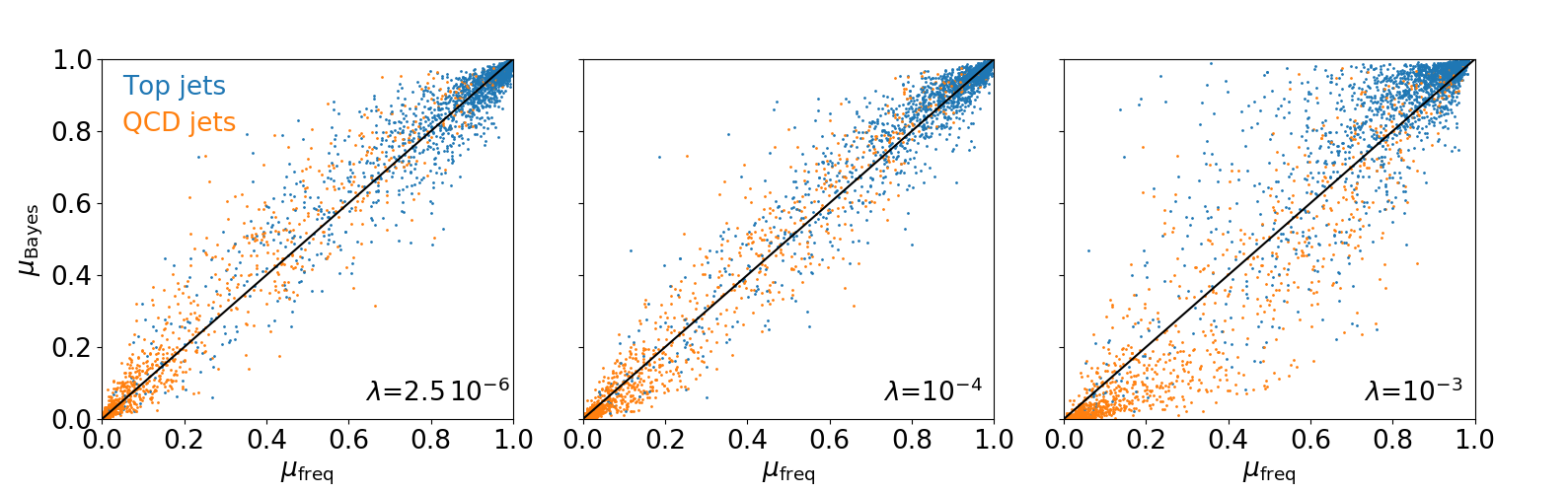}
\includegraphics[width=1.\textwidth]{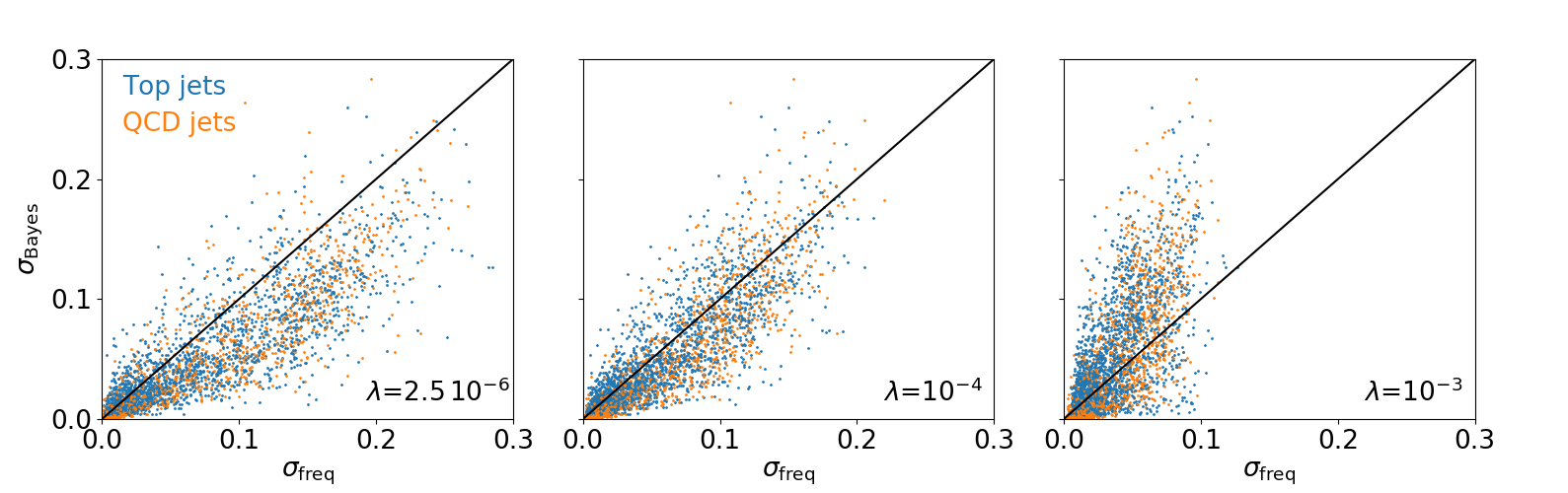}
\caption{Correlation of the mean (upper) and standard deviation
  (lower) between the BNN's $\mu_\text{pred}$ and $\sigma_\text{pred}$
  and histogrammed outputs for 100 deterministic networks. We show
  results for three different values for the L2-regularization
  $\lambda$.  Each point corresponds to one of 2000 signal and 2000
  background jets, as indicated by the colors.}
\label{fig:check_l2} 
\end{figure}
%------------------------------------------------

To investigate a frequentist interpretation of the BNN output we will
compare the predictive mean and standard deviations with the
corresponding distributions for a large number of independently
trained and tested deterministic networks.  The architecture of the
deterministic model will have the same number of layers and nodes as
the BNN.  In the deterministic network we minimize the negative
logarithm of the likelihood defined in Eq.\eqref{eq:bayes2} or,
equivalently, the cross-entropy.  Applying an additional
L2-regularization allows us to drive the network to more stable or
less complex models,
\begin{align}
L = - \log p(C|\omega) - \lambda |\omega|^2  
  \equiv \frac{\chi^2}{2} - \lambda |\omega|^2 \; ,
\end{align}
where we give the relation to the $\chi^2$ in the simplest Gaussian
approximation.  In a Bayesian approach we instead minimize
Eq.\eqref{eq:maximize}, a combination of the negative expected log-likelihood
and the KL-divergence,
\begin{align}
L
 =  - \log p(C|\omega) - \frac{\mu^2}{2 \sigma_\text{prior}^2} + \cdots
\label{eq:maximize2}
\end{align}
Relating these two approaches and identifying $\omega$ in the
deterministic network with the $\mu$ in the BNN suggests that a
Gaussian prior corresponds to an L2-regularization,
\begin{align}
\lambda = \frac{1}{2 \sigma_\text{prior}^2} \; .
\label{eq:default_lambda}
\end{align}
At this point it is crucial to emphasize that the deterministic
network has this additional option of balancing the network
performance, training time, and stability with the help of the free
L2-regularization parameter $\lambda$. In contrast, for the BNN
$\sigma_\text{prior}$ is fixed by the choice of prior.

%------------------------------------------------
\begin{figure}[t]
\centering
\includegraphics[width=1.\textwidth]{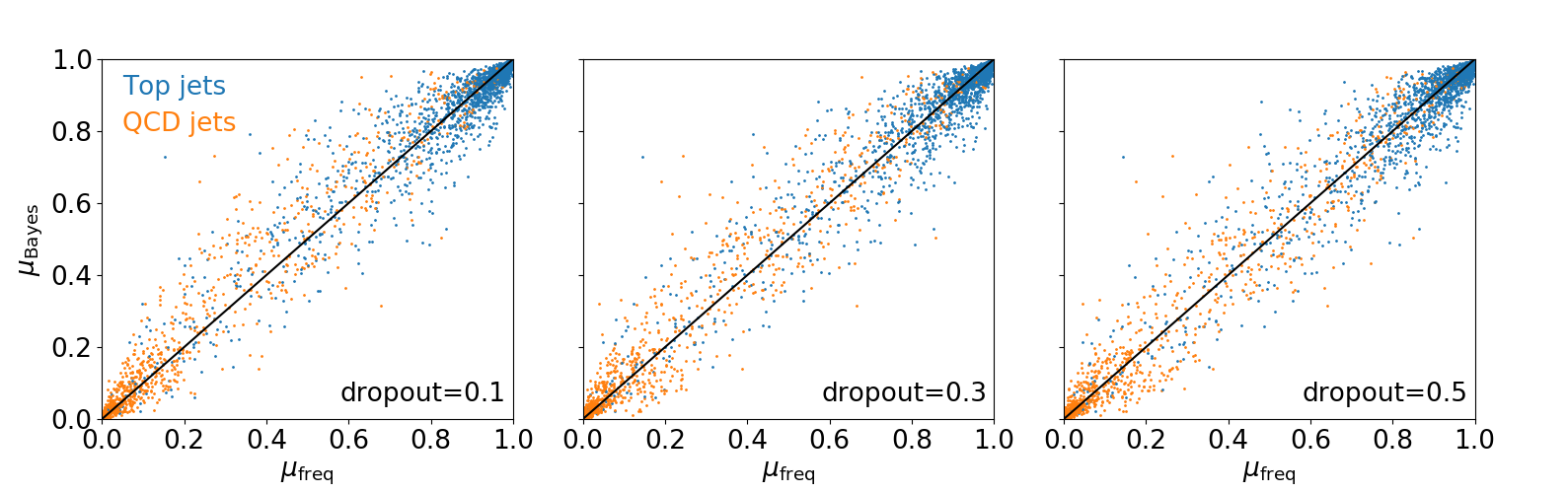}
\includegraphics[width=1.\textwidth]{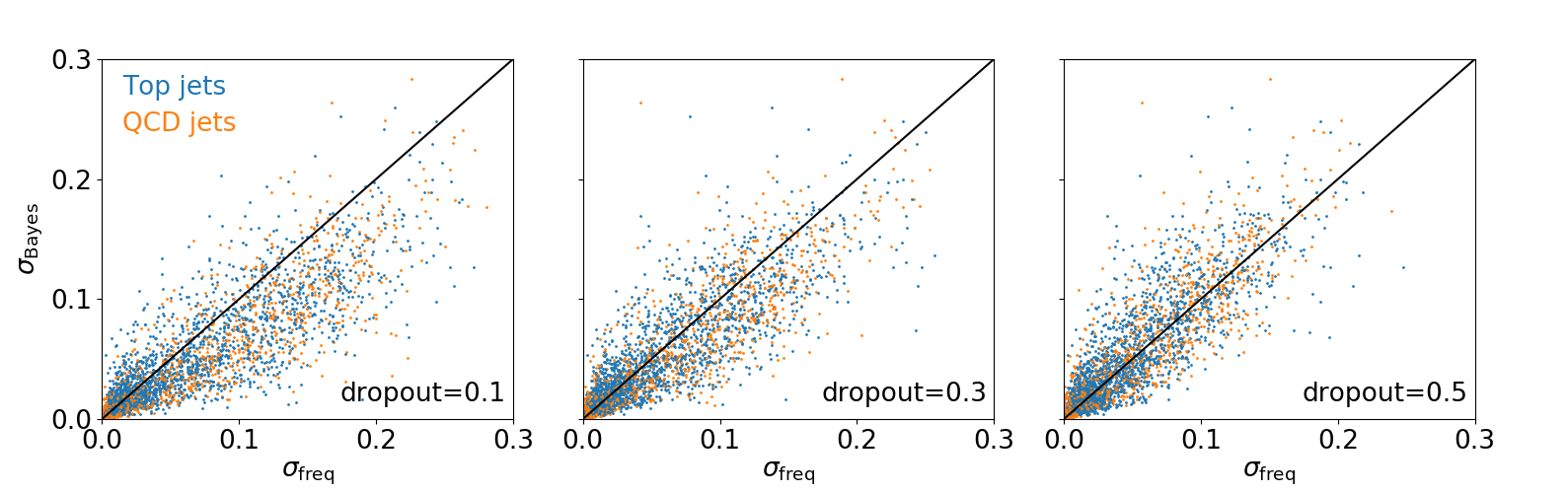}
\caption{Correlation of the mean (upper) and standard deviation
  (lower) between the BNN's $\mu_\text{pred}$ and $\sigma_\text{pred}$
  and histogrammed outputs for 80 deterministic networks with ${\lambda
  = 2.5 \cdot 10^{-6}}$. We show results for three values for the
  dropout rate.  Each point corresponds to one of 2000 signal and 2000
  background jets, as indicated by the colors.}
\label{fig:check_dropout} 
\end{figure}
%------------------------------------------------

For this comparison, we require considerably larger statistics than is
available in the public dataset based on Ref.~\cite{lola}. We generate
an new data set in the same phase-space region as in
Sec.~\ref{sec:classification}, but with 9.5M~top jets and 9.5M~QCD
jets.  We then use a deterministic toy tagger with the same
architecture as the toy BNN, but without the Bayesian features. We
construct 100 different networks training on statistically independent
samples and histogram their classification output for a given
jet. From this histogram we can extract the mean and the standard
deviation in a frequentist sense. Because the Bayesian and
deterministic networks have the same setup, we expect this
distribution of 100 means to follow the prior-independent probability
distribution from the BNN.  In Fig.~\ref{fig:check_l2} we show the
correlation between the predictive mean and predictive significance
with the results from a set of deterministic taggers. Each point in
the scatter plot corresponds to one of the 4000 jets. Jet by jet we
see that the Bayesian and frequentist mean values are clearly fully
correlated, with an increased spread for relatively poorly determined
values around $\mu = 0.5$. This spread is expected, and we have
checked that it is within the range of the predictive standard
deviation. The slight asymmetry of the taggers on the pure top and
QCD sides is not an unexpected feature.  The default value for the
deterministic L2-regularization given in Eq.\eqref{eq:default_lambda}
in our case is
\begin{align}
\lambda = 2.5 \cdot 10^{-6} \; ,
\end{align}
shown in the left panel of Fig.~\ref{fig:check_l2}. We also show how
much larger values of $\lambda$ lead to a significant underestimate of
the uncertainties in the frequentist approach, corresponding to a
strong prior-like behavior of the L2-regularization. The apparent
agreement of the two approaches not only in the mean but also in the
width is best for $\lambda = 10^{-4}$, while for the default value of
$\lambda$ the deterministic network give a slightly more conservative
error band.\bigskip

The discussion above leads us to another freedom we have in defining
our deterministic tagger: while we have not used it till now, dropout
reduces the number of neurons in each training epoch statistically to
a given percentage and is used to avoid
over-fitting~\cite{dropout}. If a neuron is switched off for one
iteration, the other neurons will compensate for this loss, leading to
a stochastic fluctuation within the network from iteration to
iteration. The key aspect of dropout is that it generates statistical
fluctuations in the network weights, and it can be shown that networks
trained with dropout are BNNs~\cite{dropout_equiv}\footnote{Strictly
  speaking, all networks using dropout are Bayesian networks, but not
  all Bayesian networks can be modelled based on dropout.}. In the
absence of a quantitative relation like Eq.\eqref{eq:default_lambda}
we have to test the dependence of our comparison between the BNN and
the deterministic networks on the dropout rate of the deterministic
networks. For the deterministic networks dropout is used during training but not during testing. In Fig.~\ref{fig:check_dropout} we show three different
correlations for dropout rates of 0.1, 0.3, and 0.5, for the default
L2-regularization.  Again, we see that small dropout leads to a
slightly larger frequentist estimate of the uncertainties, while large
dropout has a prior-like effect of reducing the frequentist error
estimate. In the center panel of we see that
\begin{align}
\lambda = 2.5 \cdot 10^{-6}
\qquad \text{and dropout rate 0.3} 
\end{align}
lead to an excellent agreement of the jet-by-jet BNN output and the
frequentist analysis of 100 deterministic networks. In practice, the
difference between these two approaches is that training a large
number of deterministic networks on an actual tagging setup is
extremely GPU-intensive, while the BNN provides $\sigma_\text{pred}$
on a jet-by-jet basis automatically with the probabilistic mean
$\mu_\text{pred}$.

Aside from the fact that we can reproduce the probability distribution
of tagging output in a frequentist sense there is another, more
conceptual lesson to learn from the this section: while the loss
function of Eq.\eqref{eq:maximize} or~\eqref{eq:maximize2} comes from
Bayes' theorem, the appearance of the KL-divergence is closely linked
to known numerical improvements of standard deterministic networks,
like dropout and L2-regularization. In that sense, the BNN with its
jet-by-jet distribution of tagging outputs is not any more Bayesian
than many of the NN-taggers which we already use.

%%%%%%%%%%%%%%%%%%%%%%%%%%%%%%%%%%%%%%%%%%%%%%%%%%%%%%%%%%%%%%%%%%%%%%%
\section{BNN top taggers}
\label{sec:top}

After understanding the concept behind BNNs and illustrating many of
their features, we are now ready to apply them to an actual physics
problem. Top tagging is an established, experimentally and
theoretically well-defined
task~\cite{top_review,jets_comparison}. Experimentally, it has the
great advantage that we can train neural networks on mixed top pair
events, where we first identify the leptonically decaying top quarks
and then run the tagger on the hadronic recoil. Theoretically, the
main features of top decays are safely perturbative~\cite{lola}, so
unlike, for instance, in the case of quark-gluon
separation~\cite{lola_qg} we do not expect detector effects and
pile-up to have decisive impact on the tagging performance. On the
other hand, detector effects and (related) systematic effects are
going to be key factors in any application of machine learning
techniques to subjet physics~\cite{aussies}, especially if we
eventually need to go beyond perfectly labelled actual jets to
MC-enhanced training samples.

%%%%%%%%%%%%%%%%%%%%%%%%%%%%%%%%%%%%%%%%%%%%%%%%%%%%%%%%%%%%%%%%%%%%%%%
\subsection{Performance}
\label{sec:benchmarking}

%------------------------------------------------
\begin{figure}[t]
\centering
\includegraphics[width=0.5\textwidth]{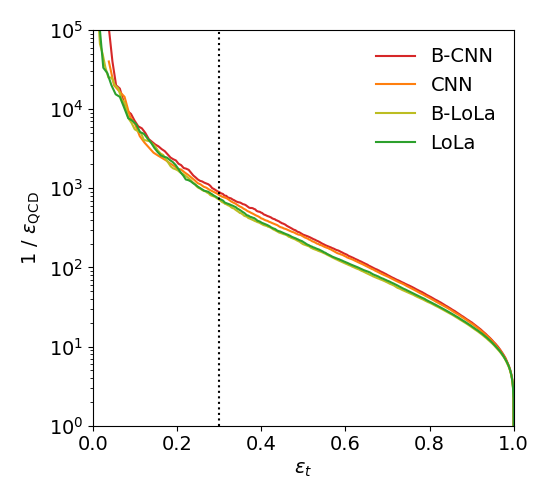}
\caption{ROC curve for the deterministic image-based and \textsc{LoLa}
  top taggers, compared to their respective BNN
  implementation.}
\label{fig:perform} 
\end{figure}
%------------------------------------------------

%------------------------------------------------
\begin{table}[b!]
\begin{center} \begin{small} \begin{tabular}{ll|cc}
\toprule
 && AUC & $1/\epsilon_\text{QCD}$ for $\epsilon_t = 30\%$ \\ 
\midrule 
CNN && 0.982 & 820 \\ 
B-CNN && 0.982 & 900 \\ 
\midrule 
LoLa & \multirow{2}{*}{$N_\text{const}=40$} & 0.979 &  630 \\
B-Lola &                                    & 0.979 & 600 \\ 
\midrule 
LoLa & \multirow{2}{*}{$N_\text{const}=100$} & 0.981 &  740 \\
B-Lola &                                     & 0.980 & 710 \\ 
\midrule 
LoLa & \multirow{2}{*}{$N_\text{const}=200$} & 0.980 & 600 \\ 
B-Lola &                                     & 0.980 & 710 \\ 
\bottomrule 
\end{tabular} \end{small} \end{center}
\caption{Performance of the different tagging architectures and their
  BNN versions on our standard top sample. }
\label{tab:perform}
\end{table}
%------------------------------------------------

To test how BNNs with different architectures react to detector
effects or systematic uncertainties, we implement BNN versions of an
image-based \textsc{DeepTop} tagger~\cite{deep_top1,deep_top2} and the
4-vector-based \textsc{DeepTopLoLa} tagger~\cite{lola}. While those
two specific implementations do not show the leading performance in
top tagging~\cite{jets_comparison}, they represent their respective
architectures with a good compromise between performance and run time.

Our test sample is the same top and QCD data set with 200k
jets~\cite{lola} as described in Sec.~\ref{sec:classification}.
Again, the jets fulfill
\begin{align}
p_{T,j} = 550~...~650~\gev
\qquad \text{and} \qquad 
|\eta_j|<2 \; .
\end{align}
The top jets are truth-matched, and the images include the improved
pre-processing taken from Ref.~\cite{deep_top2}.  The constituents for
the $\textsc{LoLa}$ tagger are extracted through the Delphes
energy-flow algorithm, and the 4-momenta of the leading 200
constituents are stored. For jets with less than 200 constituents we
simply add zero-vectors.

In Tab.~\ref{tab:perform} we show the performance of the different
deterministic taggers~\cite{lola,deep_top1,deep_top2}, as well as of
their BNN counterparts. For this application we vary the number of
ordered particle flow constituents considered by the \textsc{LoLa}
tagger between 40 and 200.  For example in Fig.~2 in Ref.~\cite{lola}
we see that the number of constituents in top jets is around 60, with
a sizeable tail towards significantly larger numbers. On the other
hand, we also know that many of them correspond to soft activity,
which according to QCD factorization is universal and essentially adds
noise. This theoretical bias is confirmed by the performances shown in
Tab~\ref{tab:perform}, where the \textsc{LoLa} tagging performance
indicates no significant improvement once we increase $N_\text{const}$
beyond 40.

In Fig.~\ref{fig:perform} we see the same behavior when we consider
the entire ROC curve for top tagging in the presence of a QCD
background. Within uncertainties related to different trainings both
taggers and their BNN counterparts show essentially the same
performance. Note that this statement only holds true in the absence
of pile-up and if we ignore statistical and systematic
uncertainties. Statistical uncertainties include, for instance, the
effect of the size of the training sample, as discussed in
Sec.~\ref{sec:statistical} and Fig.~\ref{fig:mu_sd}. We do not repeat
this exercise for the actual taggers and instead focus on detector and
systematic effects. As a starting point, we discuss how systematics
and detector effects, not accounted for in the training, affect the
predictive mean and standard deviation given by the BNN. Finally, we
show how the BNN works if we use for example a known systematic
uncertainty to modify or augment the training data.

%%%%%%%%%%%%%%%%%%%%%%%%%%%%%%%%%%%%%%%%%%%%%%%%%%%%%%%%%%%%%%%%%%%%%%%
\subsection{Systematic uncertainty from energy scale}
\label{sec:energy}

%------------------------------------------------
\begin{figure}[t]
\includegraphics[width=0.32\textwidth]{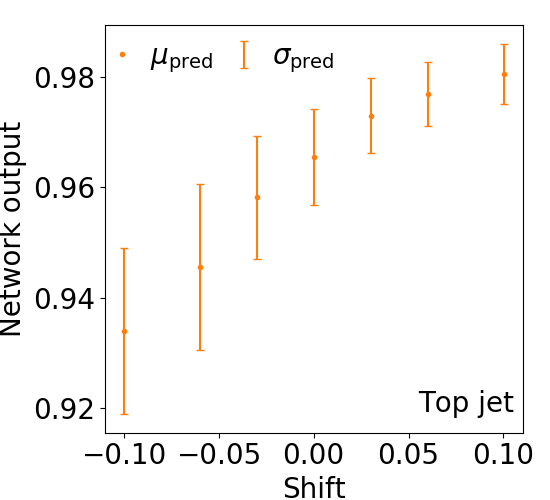} 
\includegraphics[width=0.32\textwidth]{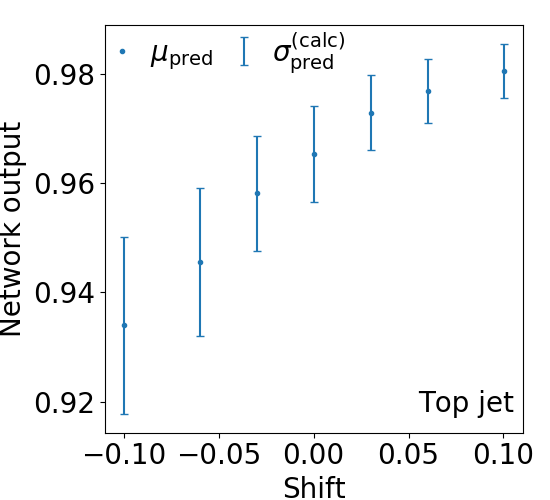}
\includegraphics[width=0.32\textwidth]{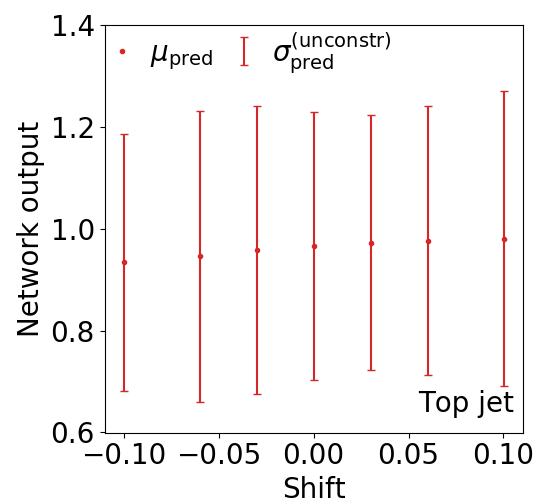}
\caption{Effect of a shifted energy scale for the hardest constituent
  on the mean and the standard deviation indicated as an error bar. We
  show the probability output (left), supplemented with the effect on
  the predictive uncertainty through the mean (center), and before the
  sigmoid transformation (right), all for the BNN \textsc{LoLa}
  tagger.}
\label{fig:systematics1}
\end{figure}
%------------------------------------------------

As part of our program of including uncertainties through BNNs, we can
investigate the jet energy uncertainty as an actual systematic
uncertainty~\cite{jes}. This means that the jet energy scale has to be
calibrated with standard candle processes, and this calibration comes
with an uncertainty from the underlying measurement and from the
extrapolation to a given event topology. We use the data set described
in Section~\ref{sec:classification} and focus on a cluster re-scaling
of the constituents of the fat jet.  It involves re-clustering the
constituents into anti-$k_T$ subjets with $R=0.4$ and in the simplest,
toy case re-scaling the energy of the leading subjet cluster. While the actual
jet energy scale uncertainty at the LHC is in the range of
$1\%~...~3\%$, we inflate the shift in this section to illustrate its
features for our networks limited by GPU time and number of training
jets.  This toy model for smearing is not meant to be experimentally
realistic, but it is chosen to identify the non-trivial features which
occur in the presence of unknown systematics.

%------------------------------------------------
\begin{figure}[t]
\centering
\includegraphics[width=0.49\textwidth]{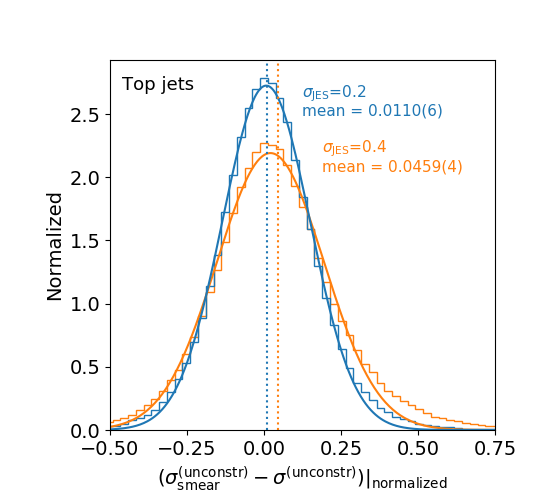}
\includegraphics[width=0.49\textwidth]{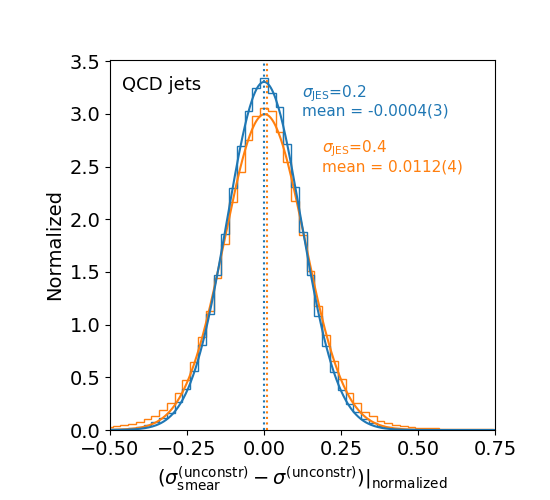}
\caption{Effect of a shifted jet energy scale on the predictive
  standard deviation output before sigmoid for the BNN \textsc{LoLa}
  tagger. We show top jets (left) and QCD jets (right) separately.}
\label{fig:systematics2}
\end{figure}
%------------------------------------------------

In a first attempt we test what happens if we train on data which does
not account for a given systematics, but we test the trained network
in the presence of a systematic shift, in this case shifting the
energy scale of the leading constituent in top and QCD jets by up to
10\%. For this study we use a BNN version of the \textsc{LoLa} tagger,
because CNN taggers usually normalize the pixel entries relative to
the total momentum.  In contrast, for the \text{LoLa} tagger we can
reduce the error bars due to statistics of the training sample to the
level where we can see systematics-induced shift in the discrimination
power and the assigned error bars. In Fig.~\ref{fig:systematics1} we
see the effect of such an energy rescaling on the network output for
one given jet. For one sign of the energy shift we see the expected
behavior, namely a loss in tagging performance combined with an
increased assigned error bar. For the other sign of the shift we see a
pattern which naturally arises whenever the systematics is not fully
de-correlated from the features the network uses to separate the two
hypotheses. In our case, we know for instance that the network will
identify jets with a more democratic energy distribution of the
constituents with QCD.  This is why a positive shift of the leading
constituent leads the network to more confidently identifying the jet
as a top jets, including a small error bar. This behavior is similar
to adversarial examples~\cite{adv_ex}, single-pixel modifications
meant to trick image recognition tools into wrong classification
obvious to human vision.

From Sec.~\ref{sec:classification} we know that the network output
after the sigmoid transformation strongly correlates the mean and
standard deviation. To understand the effect shown in the left panel
of Fig.~\ref{fig:systematics1} we therefore use
Eq.\eqref{eq:sigmoid_width_approx} to determine the shift in the
predictive standard deviation based on this correlation with the
shifted mean. In the center panel we indeed see that this correlation
dominates the change in the predictive standard deviation for this one
jet. Alternatively, we can instead show the standard deviation before
the last sigmoid layer as a function of the energy shift, confirming
this picture in the right panel.\bigskip
 
In a second step we can now look for changes in the predictive
standard deviation which are independent of the shift in the mean. For
this purpose, we train our BNN without energy smearing and test it on
jets with a smeared jet energy.  For further de-de-correlation we add a
4-vector normalization to the \textsc{LoLa tagger}, such that the
energy sum of all constituents per fat jet is one. This is a standard
step in the image-based taggers.  Unlike before, we then re-cluster
the constituents into anti-$k_T$ subjets and re-scale the energy of
each subjet (or cluster) by a random number drawn from a Gaussian
distribution with mean zero and a given standard deviation. After this
shift the predictive uncertainty given by the network before sigmoid
becomes $\sigma_\text{smear}^\text{(unconstr)}$, and we show the
distribution for the normalized shift
\begin{align}
\frac{\sigma_\text{smear}^\text{(unconstr)} - \sigma^\text{(unconstr)}}
     {(\sigma_\text{smear}^\text{(unconstr)} + \sigma^\text{(unconstr)})/2}
\end{align}
In Fig.~\ref{fig:systematics2} we show this effect separately for a
sample of top and QCD jets. For the top jets we see how for a jet
energy smearing of 20\% the predictive standard deviation increases,
but very slightly.  The fit to a symmetric Gaussian indicates how the
tail towards larger widths becomes bigger than the tail towards
smaller width. The mean of the distribution confirms a small, but
significant shift.  As a test, we also show the numbers for a
cluster-wise shift of 40\% in the jet energy scale, leading to a
consistent but larger effect. In the right panel we show the same for
QCD jets. Here, at least a 20\% jet energy re-scaling does not produce
a visible effect. The reason is that QCD jets have a comparably
democratic distribution of subjet energies, so a random smearing has
no effect of the QCD-ness in the eyes of the network unless we apply a
40\% smearing.\bigskip

%------------------------------------------------
\begin{figure}[t]
\includegraphics[width=0.49\textwidth]{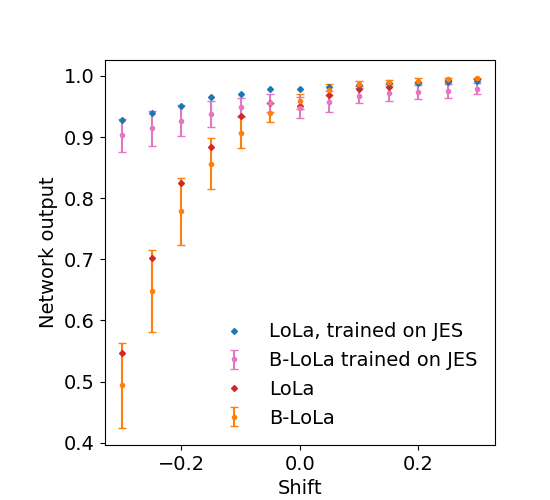}
\includegraphics[width=0.49\textwidth]{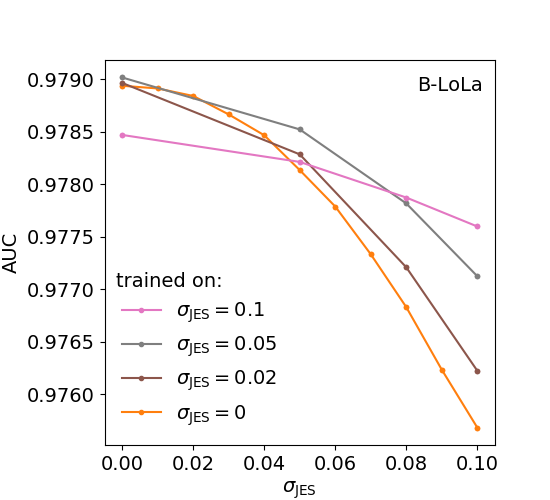}
\caption{Change of the \textsc{LoLa} tagging performance as a function
  of a jet energy smearing after training on unsmeared or augmented
  data. Left: network output for a single top jet indicating the
  predictive standard deviation as an error bar. The training sample
  with JES smearing uses $\sigma_\text{JES} = 10\%$. Right: AUC for a
  top and QCD sample for different amounts of JES smearing in the
  training sample.}
\label{fig:augment}
\end{figure}
%------------------------------------------------

Once we understand the jet energy scale systematics, we naturally want
to include them in the training. This means we will take actual
training data and manipulate it through a smearing procedure which
mimics the jet energy scale uncertainty.  We analyze the impact of
such a data augmentation for our cluster model of the jet energy scale
systematics by training the network on data after applying a Gaussian
smearing with a width of up to 10\%.  Unlike before, we now smear all
subjet clusters after re-clustering, sampling from a Gaussian. In the
left panel of Fig.~\ref{fig:augment} we first see how training with a
10\% smearing in the jet energy scale affects the tagging output on a
typical top jet. Indeed, training on augmented data stabilizes the
tagging output compared to the situation shown in
Fig.~\ref{fig:systematics1}. Specifically, we see that training on
10\% smearing pushes the drop in tagging performance to beyond that
value, while it does not affect the otherwise stable performance. This
is not a unique feature of the BNN, but also present for the
deterministic tagger.  In the right panel of Fig.~\ref{fig:augment} we
see that training on strongly smeared data will, evaluated on
un-smeared samples, lead to a decrease in performance. This is not
surprising, because smearing washes out features. However, tested on
smeared data augmented training data leads to a visible gain in
stability. While given our numerical limitation we only show this for
unrealistic shifts in the jet energy scale, the feature in itself
exists, and according to the left panel of Fig.~\ref{fig:augment} it
is significant.

%%%%%%%%%%%%%%%%%%%%%%%%%%%%%%%%%%%%%%%%%%%%%%%%%%%%%%%%%%%%%%%%%%%%%%%
\subsection{Stability tests for pile-up}
\label{sec:pileup}

%------------------------------------------------
\begin{figure}[t]
\includegraphics[scale=0.49]{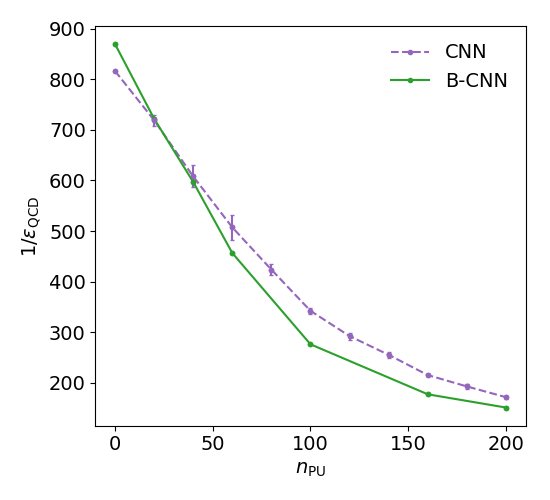}
\includegraphics[scale=0.49]{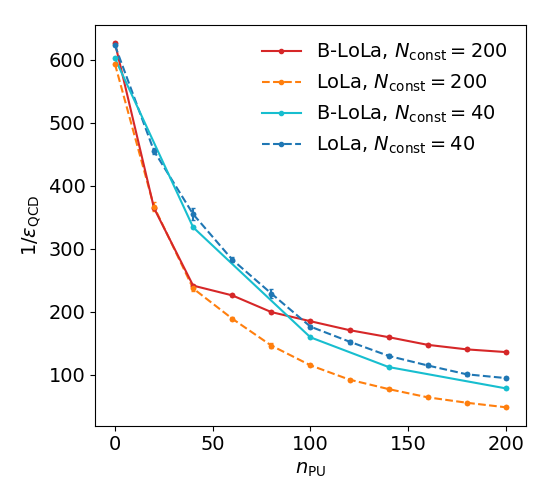}
\caption{Impact on pile-up for the image-based tagger (left) and the
  \textsc{LoLa} tagger with two numbers of constituents (right).}
\label{fig:stability_pileup}
\end{figure}
%------------------------------------------------

Finally, we can test how BNNs deal with systematic effects in the data,
which are not accounted for as an uncertainty. For instance pile-up is
known to lead to problems for subjet physics tools. Strictly speaking,
pile-up is not a systematic uncertainty, because we can measure it
statistically and even attempt to remove it event by event. It arises
from a high multiplicity of interactions per bunch crossing.  To
remove it, some standard tools such as \textsc{Puppi}~\cite{puppi} and
\textsc{SoftKiller}~\cite{softkiller} are typically employed, and more
recently there has been successful applications of machine learning
methods to this problem~\cite{MLpileup}.  In this section we test how
stable deterministic and BNN approaches are to the amount of pile-up
in an event with jets.

%------------------------------------------------
\begin{figure}[t]
\centering
\includegraphics[width=0.49\textwidth]{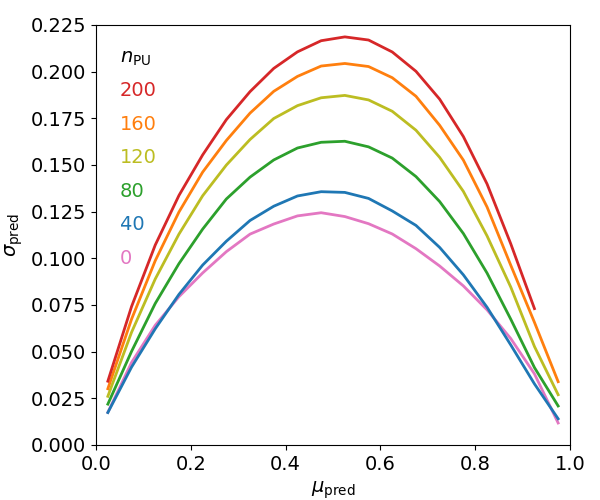}
\includegraphics[width=0.49\textwidth]{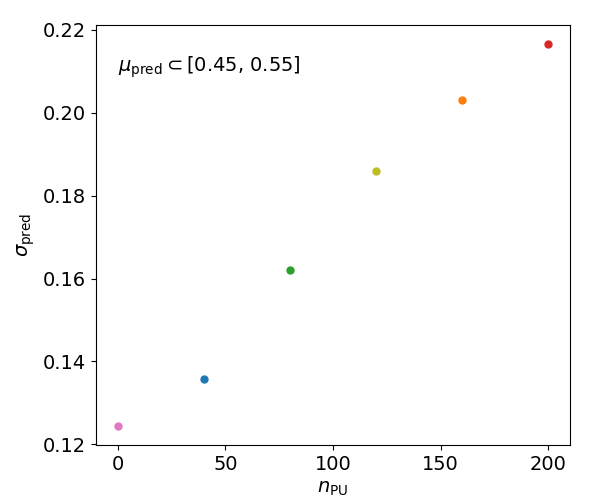}\\
\includegraphics[width=0.24\textwidth]{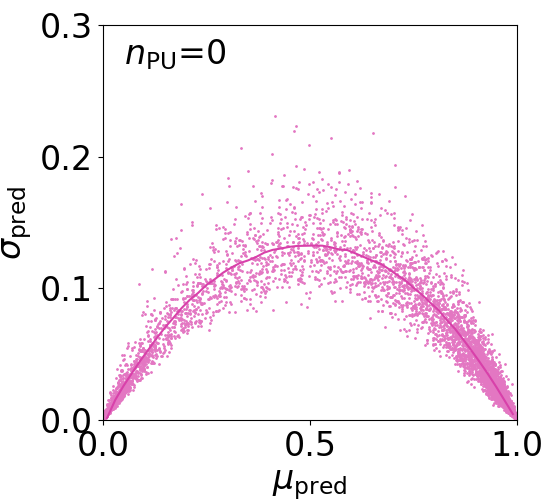}
\includegraphics[width=0.24\textwidth]{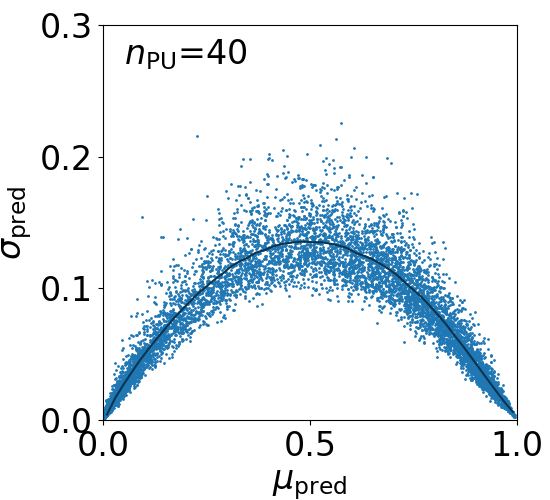}
\includegraphics[width=0.24\textwidth]{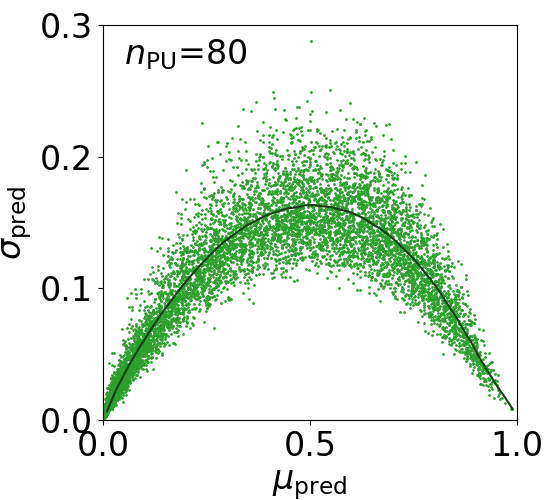}
\includegraphics[width=0.24\textwidth]{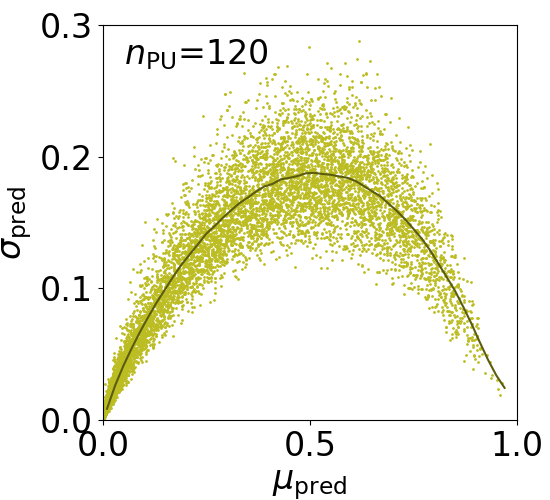}
\caption{Correlation between predictive mean and standard deviation
  for the BNN \textsc{LoLa} network with pile-up included. The right
  panel shows the predictive standard deviation for $\mu_\text{pred} =
  0.45~...~0.55$ as a function of the number of added pile-up events. The lower panels instead show the statistical
  spread for 10k jets, signal and background combined.}
\label{fig:money_pileup} 
\end{figure}
%------------------------------------------------

%------------------------------------------------
\begin{figure}[t]
\includegraphics[width=0.49\textwidth]{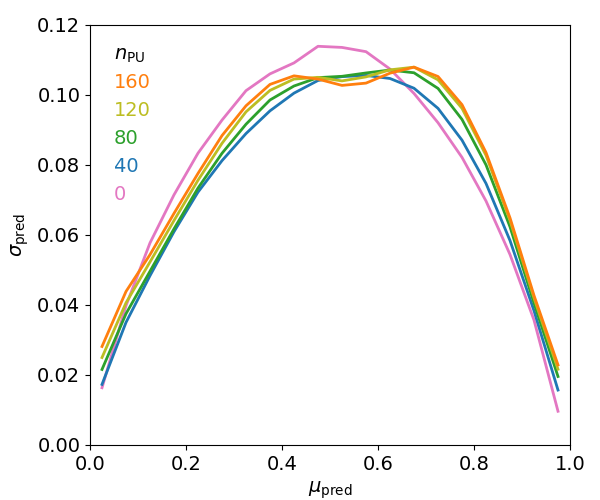}
\includegraphics[width=0.49\textwidth]{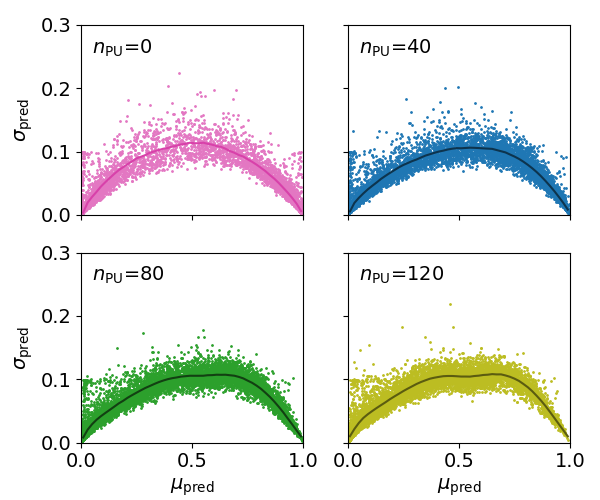}
\caption{Stability of the image-based BNN in the presence of pile-up,
  in analogy to Fig.~\ref{fig:money_pileup}.}
\label{fig:bcnn_mean_std_plane}
\end{figure}
%------------------------------------------------

To simulate pile-up, we generate min-bias events again with
\textsc{Pythia}8~\cite{pythia}, including
\textsc{Delphes}~\cite{delphes} and with the same settings as in
Ref.~\cite{deep_top1}. 
%In addition to the 4-momenta of each particle,
%we also store the information of whether or not the particle has an
%electric charge.
In total, we generate 1M min-bias events with the hardest 400
constituents of each event.  We then add a variable number of up to
200 min-bias events on top of the signal and background jet events.
For this combination we re-cluster the fat jets and select the hardest
jet per event. The new jets are pre-processed the same way as the jets
without pile-up described in Sec.~\ref{sec:benchmarking}.  For the
stability test we train the networks on a sample without pile-up and
test it on samples with different amount of pile-up. We note that this
comparison ignores the fact that we can actually simulate pile-up and
that experimental training data tends to include pile-up at some
level, so all we test is the stability of the taggers.\bigskip

In Fig.~\ref{fig:stability_pileup} we show the background rejection at
fixed signal efficiency $\epsilon_t = 30\%$ as a function of the
number of pile-up events added. While for the CNN and its BNN version
we see no significant difference in performance, the \textsc{LoLa}
tagger shows some interesting features. The reason for this different
behavior is the fact that in the latter the constituents are not just
added noise, but given in an ordered manner. As long as we only use
$N_\text{const} =40$ constituents the \textsc{LoLa} tagger simply
ignores the relatively soft contributions from pile-up. In that sense
it is by construction insensitive to universal soft activity,
independent of its source. We have confirmed this pattern for our tagger.

The picture changes when we start to include up to
$N_\text{const}=200$ constituents. In Fig.~\ref{fig:stability_pileup}
we see how the deterministic tagger is now sensitive to noise effects,
and how an increased amount of noise cuts into the tagging
performance. However, the behavior of the BNN version of the
\textsc{LoLa} tagger is different; here, the BNN setup allows the
tagger to utilize the additional information, even though there exists
only soft QCD radiation related to a single bunch crossing in the
training, and turn it into a performance improvement. In
Fig.~\ref{fig:money_pileup} we see, however, that this improvement of
the performance comes at a price in the uncertainty. For an increased
number of pile-up events the predictive standard deviation increases
from 12\% to 22\%. In the lower four panels we see that this
improvement happens in a perfectly stable system, maintaining the
expected parabolic correlation between the mean and standard deviation
outputs of the tagger.

Finally, we can apply the same test to the BNN version of the
image-based tagger.  According to Fig.~\ref{fig:stability_pileup} the
performance of the CNN taggers is relatively stable with respect to
pile-up. In Fig.~\ref{fig:bcnn_mean_std_plane} we show the performance
of the Bayesian CNN tagger in more detail and immediately see that for
pile-up values of 60-100 the network becomes unstable and violates the
parabolic shape predicted for the probabilistic output in
Eq.\eqref{eq:sigmoid_width_approx}. This parabolic correlation,
however, is a fundamental effect of the network output on the closed
interval $[0,1]$. The deviations start affecting the more poorly
classified jets and is not immediately visible for example from the
uncorrelated output of the predictive mean and the predictive standard
deviation.  This means that the careful analysis of the predictive
standard deviation allows us to gain addition insight on the stability
of the network, which we do not gain from a performance study of the
deterministic network.

%%%%%%%%%%%%%%%%%%%%%%%%%%%%%%%%%%%%%%%%%%%%%%%%%%%%%%%%%%%%%%%%%%%%%%%
\section{Outlook}

Machine learning applied to low-level detector observables has the
potential to transform many aspects of LHC analyses, for instance
subjet analyses and top tagging. Two open questions for instance of
classification networks, independent of their architecture and setup,
are how we can include a proper error treatment and how we can
understand the network output.  Bayesian neural networks offer solutions to
both of these problem, also going beyond what established tagging
approaches can offer. Their list of advantages and opportunities for
LHC applications is remarkable: (i) they can classify jets or events
including error bars; (ii) they can can be interpreted in a
frequentist sense, the prior has no visible impact; (iii) they help us
understand regularization and dropout in deterministic networks; (iv)
they are usually well-calibrated, even though re-calibration can
always streamline LHC applications. We have illustrated these features
and their application to LHC physics using a toy top tagger working on
jet images.

The standard way classifiers in HEP analyses are currently employed is to determine a working point and then assess the uncertainty on its signal efficiency and background rejection. A classifier with per-jet or per-event uncertainties automatically provides these as well but with additional information provided by the uncertainty which could be potentially included in a statistical analysis.
 We have shown that standard deep-learning tagging frameworks, like an
image-based CNN tagger or a 4-vector-based \textsc{LoLa} tagger, can
be easily extended to a BNN version. Specifically, we have shown how
the Bayesian versions 
\begin{itemize}
\setlength\itemsep{0em}
\item[$\cdot$] track the statistical uncertainty from a limited training
  sample, Fig.~\ref{fig:mu_sd};
\item[$\cdot$] track mean-correlated systematics from the jet energy scale,
  Fig.~\ref{fig:systematics1};
\item[$\cdot$] track systematics orthogonal to the correlation with the mean,
  Fig.~\ref{fig:systematics2};
\item[$\cdot$] provide a handle on the stability with respect to pile-up,
  Fig.~\ref{fig:stability_pileup};
\item[$\cdot$] have the same performance as deterministic networks,
  Fig.~\ref{fig:perform}.
\end{itemize}
We note that all the above advantages of the BNN extensions are
available at essentially no cost, but they provide a wealth of
additional information and insight into the behavior of deep
networks. Especially for notorious users in particle physics these
additional handles controlling and understanding the network output
should be very attractive. They might also be useful in determining
the best-suited classification network architectures for ATLAS and
CMS.

\bigskip
%%%%%%%%%%%%%%%%%%%%%%%%%%%%%%%%%%%%%%%%%%%%%%%%%%%%%%%%%%%%%%%%%%%%%%
\begin{center} \textbf{Acknowledgments} \end{center}

We would like to thank Fred Hamprecht and Ullrich K\"othe for
brokering this collaboration and for their support. We could also like
to thank Ben Nachman for some very useful discussions on deep learning
and uncertainties, and Kyle Cranmer for pointing us to some early
papers.  The authors acknowledge support by the state of
Baden-W\"urttemberg through bwHPC and the German Research Foundation
(DFG) through grant no INST 39/963-1 FUGG. JT would like to thank BMBF
for funding.

%%%%%%%%%%%%%%%%%%%%%%%%%%%%%%%%%%%%%%%%%%%%%%%%%%%%%%%%%%%%%%%%%%%%%%


\begin{thebibliography}{99}

\bibitem{early_stuff}
  L.~Lonnblad, C.~Peterson and T.~Rognvaldsson,
  %``Using neural networks to identify jets,''
  Nucl.\ Phys.\ B {\bf 349}, 675 (1991)
  \doi{10.1016/0550-3213(91)90392-B}.
  %%CITATION = doi:10.1016/0550-3213(91)90392-B;%%
  %105 citations counted in INSPIRE as of 16 Nov 2018

\bibitem{jet_images}
  J.~Cogan, M.~Kagan, E.~Strauss and A.~Schwarztman,
  %``Jet-Images: Computer Vision Inspired Techniques for Jet Tagging,''
  JHEP {\bf 1502}, 118 (2015)
  \doi{10.1007/JHEP02(2015)118}
  [\arxiv{1407.5675} [hep-ph]];
  %%CITATION = doi:10.1007/JHEP02(2015)118;%%
  %13 citations counted in INSPIRE as of 06 Jan 2017
  L.~de Oliveira, M.~Kagan, L.~Mackey, B.~Nachman and A.~Schwartzman,
  %``Jet-images — deep learning edition,''
  JHEP {\bf 1607}, 069 (2016)
  \doi{10.1007/JHEP07(2016)069}
  [\arxiv{1511.05190} [hep-ph]].
  %%CITATION = doi:10.1007/JHEP07(2016)069;%%
  %12 citations counted in INSPIRE as of 06 Jan 2017

\bibitem{particlenet} 
  H.~Qu and L.~Gouskos,
  %``ParticleNet: Jet Tagging via Particle Clouds,''
  \arxiv{1902.08570} [hep-ph].
  %%CITATION = ARXIV:1902.08570;%%

\bibitem{jets_qg}
  J.~Gallicchio and M.~D.~Schwartz,
  %``Quark and Gluon Jet Substructure,''
  JHEP {\bf 1304}, 090 (2013)
  \doi{10.1007/JHEP04(2013)090}
  [\arxiv{1211.7038} [hep-ph]];
  %%CITATION = doi:10.1007/JHEP04(2013)090;%%
  %67 citations counted in INSPIRE as of 16 Nov 2018
  P.~T.~Komiske, E.~M.~Metodiev and M.~D.~Schwartz,
  %``Deep learning in color: towards automated quark/gluon jet discrimination,''
  JHEP {\bf 1701}, 110 (2017)
  \doi{10.1007/JHEP01(2017)110}
  [\arxiv{1612.01551} [hep-ph]];
  %%CITATION = doi:10.1007/JHEP01(2017)110;%%
  %38 citations counted in INSPIRE as of 06 Apr 2018
  T.~Cheng,
  %``Recursive Neural Networks in Quark/Gluon Tagging,''
  Comput.\ Softw.\ Big Sci.\  {\bf 2}, no. 1, 3 (2018)
  \doi{10.1007/s41781-018-0007-y}
  [\arxiv{1711.02633} [hep-ph]];
  %%CITATION = doi:10.1007/s41781-018-0007-y;%%
  %11 citations counted in INSPIRE as of 06 Aug 2018
  P.~T.~Komiske, E.~M.~Metodiev and J.~Thaler,
  %``Energy Flow Networks: Deep Sets for Particle Jets,''
  \arxiv{1810.05165} [hep-ph];
  %%CITATION = ARXIV:1810.05165;%%
  %1 citations counted in INSPIRE as of 05 Dec 2018
  S.~Bright-Thonney and B.~Nachman,
  %``Investigating the Topology Dependence of Quark and Gluon Jets,''
  \arxiv{1810.05653} [hep-ph].
  %%CITATION = ARXIV:1810.05653;%%

\bibitem{lola_qg} 
  G.~Kasieczka, N.~Kiefer, T.~Plehn and J.~M.~Thompson,
  %``Quark-Gluon Tagging: Machine Learning meets Reality,''
  arXiv:1812.09223 [hep-ph].
  %%CITATION = ARXIV:1812.09223;%%

\bibitem{jets_w}
  P.~Baldi, K.~Bauer, C.~Eng, P.~Sadowski and D.~Whiteson,
  %``Jet Substructure Classification in High-Energy Physics with Deep Neural Networks,''
  Phys.\ Rev.\ D {\bf 93}, no. 9, 094034 (2016)
  \doi{10.1103/PhysRevD.93.094034}
  [\arxiv{1603.09349} [hep-ex]];
  %%CITATION = doi:10.1103/PhysRevD.93.094034;%%
  %6 citations counted in INSPIRE as of 06 Jan 2017
  G.~Louppe, K.~Cho, C.~Becot and K.~Cranmer,
  %``QCD-Aware Recursive Neural Networks for Jet Physics,''
  JHEP {\bf 1901}, 057 (2019)
  \doi{10.1007/JHEP01(2019)057}
  [\arxiv{1702.00748} [hep-ph]].
  %%CITATION = doi:10.1007/JHEP01(2019)057;%%
  %47 citations counted in INSPIRE as of 24 Jan 2019

\bibitem{jets_h}
  S.~H.~Lim and M.~M.~Nojiri,
  %``Spectral Analysis of Jet Substructure with Neural Network: Boosted Higgs Case,''
  \arxiv{1807.03312} [hep-ph];
  %%CITATION = ARXIV:1807.03312;%%
  %2 citations counted in INSPIRE as of 12 Oct 2018
  J.~Lin, M.~Freytsis, I.~Moult and B.~Nachman,
  %``Boosting $H\to b\bar b$ with Machine Learning,''
  \arxiv{1807.10768} [hep-ph].
  %%CITATION = ARXIV:1807.10768;%%

\bibitem{deep_top1}
  G.~Kasieczka, T.~Plehn, M.~Russell and T.~Schell,
  %``Deep-learning Top Taggers or The End of QCD?,''
  JHEP {\bf 1705}, 006 (2017)
  \doi{10.1007/JHEP05(2017)006}
  [\arxiv{1701.08784} [hep-ph]].
  %%CITATION = doi:10.1007/JHEP05(2017)006;%%
  %4 citations counted in INSPIRE as of 17 Jun 2017
   
\bibitem{deep_top2} 
  S.~Macaluso and D.~Shih,
  %``Pulling Out All the Tops with Computer Vision and Deep Learning,''
  JHEP {\bf 1810}, 121 (2018)
  \doi{10.1007/JHEP10(2018)121}
  [\arxiv{1803.00107} [hep-ph]].
  %%CITATION = doi:10.1007/JHEP10(2018)121;%%
  %21 citations counted in INSPIRE as of 01 Feb 2019

\bibitem{lola} 
  A.~Butter, G.~Kasieczka, T.~Plehn and M.~Russell,
  %``Deep-learned Top Tagging with a Lorentz Layer,''
  SciPost Phys.\  {\bf 5}, 028 (2018)
  \doi{10.21468/SciPostPhys.5.3.028}
  [\arxiv{1707.08966} [hep-ph]].
  %%CITATION = doi:10.21468/SciPostPhys.5.3.028;%%
  %29 citations counted in INSPIRE as of 12 Oct 2018

\bibitem{jets_top}
  L.~G.~Almeida, M.~Backovic, M.~Cliche, S.~J.~Lee and M.~Perelstein,
  %``Playing Tag with ANN: Boosted Top Identification with Pattern Recognition,''
  JHEP {\bf 1507}, 086 (2015)
  \doi{10.1007/JHEP07(2015)086}
  [\arxiv{1501.05968} [hep-ph]];
  %%CITATION = doi:10.1007/JHEP07(2015)086;%%
  %13 citations counted in INSPIRE as of 06 Jan 2017
  J.~Pearkes, W.~Fedorko, A.~Lister and C.~Gay,
  %``Jet Constituents for Deep Neural Network Based Top Quark Tagging,''
  \arxiv{1704.02124} [hep-ex];
  %%CITATION = ARXIV:1704.02124;%%
  %2 citations counted in INSPIRE as of 23 Jul 2017
  S.~Egan, W.~Fedorko, A.~Lister, J.~Pearkes and C.~Gay,
  %``Long Short-Term Memory (LSTM) networks with jet constituents for boosted top tagging at the LHC,''
  \arxiv{1711.09059} [hep-ex];
  %%CITATION = ARXIV:1711.09059;%%
  %10 citations counted in INSPIRE as of 06 Aug 2018  
  S.~Choi, S.~J.~Lee and M.~Perelstein,
  %``Infrared Safety of a Neural-Net Top Tagging Algorithm,''
  \arxiv{1806.01263} [hep-ph];
  %%CITATION = ARXIV:1806.01263;%%
  %2 citations counted in INSPIRE as of 06 Aug 2018  
  L.~Moore, K.~Nordstr\"om, S.~Varma and M.~Fairbairn,
  %``Reports of My Demise Are Greatly Exaggerated: $N$-subjettiness Taggers Take On Jet Images,''
  \arxiv{1807.04769} [hep-ph].
  %%CITATION = ARXIV:1807.04769;%%
  %1 citations counted in INSPIRE as of 06 Aug 2018  

\bibitem{ml_review}
  for a review see \eg
  A.~J.~Larkoski, I.~Moult and B.~Nachman,
  %``Jet Substructure at the Large Hadron Collider: A Review of Recent Advances in Theory and Machine Learning,''
  \arxiv{1709.04464} [hep-ph].
  %%CITATION = ARXIV:1709.04464;%%
  %53 citations counted in INSPIRE as of 06 Aug 2018

\bibitem{jets_comparison}
  G.~Kasieczka {\it et al.},
  %``The Machine Learning Landscape of Top Taggers,''
  \arxiv{1902.09914} [hep-ph].
  %%CITATION = ARXIV:1902.09914;%%

\bibitem{aussies}
  J.~Barnard, E.~N.~Dawe, M.~J.~Dolan and N.~Rajcic,
  %``Parton Shower Uncertainties in Jet Substructure Analyses with Deep Neural Networks,''
  Phys.\ Rev.\ D {\bf 95}, no. 1, 014018 (2017)
  \doi{10.1103/PhysRevD.95.014018}
  [\arxiv{1609.00607} [hep-ph]].
  %%CITATION = ARXIV:1609.00607;%%
  %2 citations counted in INSPIRE as of 06 Jan 2017

\bibitem{bad}
  C.~Englert, P.~Galler, P.~Harris and M.~Spannowsky,
  %``Machine Learning Uncertainties with Adversarial Neural Networks,''
  Eur.\ Phys.\ J.\ C {\bf 79}, no. 1, 4 (2019)
  \doi{10.1140/epjc/s10052-018-6511-8}
  [\arxiv{1807.08763} [hep-ph]].
  %%CITATION = doi:10.1140/epjc/s10052-018-6511-8;%%
  %6 citations counted in INSPIRE as of 18 Apr 2019

\bibitem{weak}
  L.~M.~Dery, B.~Nachman, F.~Rubbo and A.~Schwartzman,
  %``Weakly Supervised Classification in High Energy Physics,''
  JHEP {\bf 1705}, 145 (2017)
  \doi{10.1007/JHEP05(2017)145}
  [\arxiv{1702.00414} [hep-ph]];
  %%CITATION = doi:10.1007/JHEP05(2017)145;%%
  %18 citations counted in INSPIRE as of 30 Jul 2018
  T.~Cohen, M.~Freytsis and B.~Ostdiek,
  %``(Machine) Learning to Do More with Less,''
  JHEP {\bf 1802}, 034 (2018)
  \doi{10.1007/JHEP02(2018)034}
  [\arxiv{1706.09451} [hep-ph]];
  %%CITATION = doi:10.1007/JHEP02(2018)034;%%
  %22 citations counted in INSPIRE as of 30 Jul 2018
  E.~M.~Metodiev, B.~Nachman and J.~Thaler,
  %``Classification without labels: Learning from mixed samples in high energy physics,''
  JHEP {\bf 1710}, 174 (2017)
  \doi{10.1007/JHEP10(2017)174}
  [\arxiv{1708.02949} [hep-ph]].
  %%CITATION = doi:10.1007/JHEP10(2017)174;%%
  %26 citations counted in INSPIRE as of 30 Jul 2018
  P.~T.~Komiske, E.~M.~Metodiev, B.~Nachman and M.~D.~Schwartz,
  %``Learning to classify from impure samples with high-dimensional data,''
  Phys.\ Rev.\ D {\bf 98}, no. 1, 011502 (2018)
  \doi{10.1103/PhysRevD.98.011502}
  [\arxiv{1801.10158} [hep-ph]];
  %%CITATION = doi:10.1103/PhysRevD.98.011502;%%
  %11 citations counted in INSPIRE as of 30 Jul 2018
  A.~Andreassen, I.~Feige, C.~Frye and M.~D.~Schwartz,
  %``JUNIPR: a Framework for Unsupervised Machine Learning in Particle Physics,''
  \arxiv{1804.09720} [hep-ph];
  %%CITATION = ARXIV:1804.09720;%%
  %4 citations counted in INSPIRE as of 18 Jul 2018
  T.~Heimel, G.~Kasieczka, T.~Plehn and J.~M.~Thompson,
  %``QCD or What?,''
  \arxiv{1808.08979} [hep-ph];
  %%CITATION = ARXIV:1808.08979;%%
  %7 citations counted in INSPIRE as of 25 Feb 2019
  M.~Farina, Y.~Nakai and D.~Shih,
  %``Searching for New Physics with Deep Autoencoders,''
  \arxiv{1808.08992} [hep-ph].
  %%CITATION = ARXIV:1808.08992;%%
  %6 citations counted in INSPIRE as of 25 Feb 2019
  O.~Cerri, T.~Q.~Nguyen, M.~Pierini, M.~Spiropulu and J.~R.~Vlimant,
  %``Variational Autoencoders for New Physics Mining at the Large Hadron Collider,''
  \arxiv{1811.10276} [hep-ex].
  %%CITATION = ARXIV:1811.10276;%%
  %1 citations counted in INSPIRE as of 25 Feb 2019

\bibitem{information} 
  K.~Datta and A.~Larkoski,
  %``How Much Information is in a Jet?,''
  JHEP {\bf 1706}, 073 (2017)
  \doi{10.1007/JHEP06(2017)073}
  [\arxiv{1704.08249} [hep-ph]];
  %%CITATION = \doi{10.1007/JHEP06(2017)073;%%
  %21 citations counted in INSPIRE as of 06 Aug 2018  
  S.~Chang, T.~Cohen and B.~Ostdiek,
  %``What is the Machine Learning?,''
  Phys.\ Rev.\ D {\bf 97}, no. 5, 056009 (2018)
  \doi{10.1103/PhysRevD.97.056009}
  [\arxiv{1709.10106} [hep-ph]];
  %%CITATION = doi:10.1103/PhysRevD.97.056009;%%
  %12 citations counted in INSPIRE as of 12 Oct 2018
  K.~Datta and A.~J.~Larkoski,
  %``Novel Jet Observables from Machine Learning,''
  JHEP {\bf 1803}, 086 (2018)
  \doi{10.1007/JHEP03(2018)086}
  [\arxiv{1710.01305} [hep-ph]];
  %%CITATION = doi:10.1007/JHEP03(2018)086;%%
  %16 citations counted in INSPIRE as of 12 Oct 2018
  P.~T.~Komiske, E.~M.~Metodiev and J.~Thaler,
  %``Energy flow polynomials: A complete linear basis for jet substructure,''
  JHEP {\bf 1804}, 013 (2018)
  \doi{10.1007/JHEP04(2018)013}
  [\arxiv{1712.07124} [hep-ph]];
  %%CITATION = doi:10.1007/JHEP04(2018)013;%%
  %20 citations counted in INSPIRE as of 11 Feb 2019
  T.~Roxlo and M.~Reece,
  %``Opening the black box of neural nets: case studies in stop/top discrimination,''
  \arxiv{1804.09278} [hep-ph];
  %%CITATION = ARXIV:1804.09278;%%
  %4 citations counted in INSPIRE as of 12 Oct 2018

\bibitem{qjets}
  for some ideas on uncertainties in jet classification see \eg
  S.~D.~Ellis, A.~Hornig, D.~Krohn and T.~S.~Roy,
  %``On Statistical Aspects of Qjets,''
  JHEP {\bf 1501}, 022 (2015)
  \doi{10.1007/JHEP01(2015)022}
  [\arxiv{1409.6785} [hep-ph]];
  %%CITATION = doi:10.1007/JHEP01(2015)022;%%
  %8 citations counted in INSPIRE as of 06 Aug 2019
  L.~Mackey, B.~Nachman, A.~Schwartzman and C.~Stansbury,
  %``Fuzzy Jets,''
  JHEP {\bf 1606}, 010 (2016)
  \doi{10.1007/JHEP06(2016)010}
  [\arxiv{1509.02216} [hep-ph]].
  %%CITATION = doi:10.1007/JHEP06(2016)010;%%
  %4 citations counted in INSPIRE as of 06 Aug 2019

\bibitem{medical}
  for a similar study in a different field see \eg 
  C~Leibig, V~Allken, M.~S.~Ayhan1, P.~Berens, and S.~Wahl,
  %`Leveraging uncertainty information from deep neural networks for disease detection
  Scientific Reports {\bf 7}, 17816 (2017)
  \doi{10.1038/s41598-017-17876-z}.

\bibitem{deep_errors}
  A.~Kendall and Y.~Gal, 
  %What Uncertainties Do We Need in Bayesian Deep Learning for Computer Vision?
  Proc. NIPS (2017)
  [\arxiv{1703.04977} [cs.CV]].

\bibitem{bnn_early}
  D.~MacKay 
  %`Probable Networks and Plausible Predictions – A Review of Practical Bayesian Methods for Supervised Neural Networks,''
  \href{http://www.inference.org.uk/mackay/network.pdf}{Comp. in Neural Systems, {\bf 6}, 4679 (1995)};
  R.~Neal,
  %`Bayesian learning for neural networks,''
  \href{ftp://www.cs.toronto.edu/dist/radford/thesis.pdf}{doctoral thesis}, Toronto 1995;
  Y.~Gal, 
  %``Uncertainty in Deep Learning,''
  \href{http://mlg.eng.cam.ac.uk/yarin/thesis/thesis.pdf}{doctoral thesis}, Cambridge 2016.

\bibitem{bnn_tev}
  P.~C.~Bhat and H.~B.~Prosper,
  %``Bayesian neural networks,''
  \href{http://www.physics.ox.ac.uk/phystat05/proceedings/files//bhat_prosper_phystat05.pdf}{Conf.\ Proc.\ C {\bf 050912}, 151 (2005)};
  %%CITATION = CONFP,C050912,151;%%
  %4 citations counted in INSPIRE as of 06 Aug 2019
  S.~R.~Saucedo,
  %``Bayesian Neural Networks for Classification,''
  Florida State University, 
  \href{http://purl.flvc.org/fsu/fd/FSU_migr_etd-2069}{master thesis (2007)}

\bibitem{bnn_nu}
  Y.~Xu, W.~w.~Xu, Y.~x.~Meng, K.~Zhu and W.~Xu,
  %``Applying Bayesian Neural Networks to Event Reconstruction in Reactor Neutrino Experiments,''
  Nucl.\ Instrum.\ Meth.\ A {\bf 592}, 451 (2008)
  \doi{10.1016/j.nima.2008.04.006}
  [\arxiv{0712.4042} [physics.data-an]].
  %%CITATION = doi:10.1016/j.nima.2008.04.006;%%
  %4 citations counted in INSPIRE as of 06 Aug 2019

\bibitem{blei}
 D.~M.~Blei,
  %`Variational Inference: A Review for Statisticians 
  JASA {\bf 112}, 859 (2017)
  \doi{10.1080/01621459.2017.1285773}
  [\arxiv{1601.00670} [stat.CO]].

\bibitem{tensorflow_probability}
  M.~Abadi {\it et al.},
  %`TensorFlow: A System for Large-Scale Machine Learning',
  OSDI {\bf 16}, 265 (2016);
  %Library for probabilistic networks
 \href{https://www.tensorflow.org/probability}{Tensorflow probability}

\bibitem{flipout}
  %Method for calculating gradients
  Y.~Wen, P.~Vicol, J.~Ba, D.~Tran, and R.~Grosse (2018)
  [\arxiv{1803.04386} [cs.LG]]

\bibitem{pythia}
  T.~Sj\"ostrand {\it et al.},
  %``An Introduction to PYTHIA 8.2,''
  Comput.\ Phys.\ Commun.\  {\bf 191} (2015) 159
  \doi{10.1016/j.cpc.2015.01.024}
  [\arxiv{1410.3012} [hep-ph]].
  %%CITATION = doi:10.1016/j.cpc.2015.01.024;%%
  %1444 citations counted in INSPIRE as of 11 Feb 2019

\bibitem{delphes} 
  J.~de Favereau {\it et al.} [DELPHES 3 Collaboration],
  %``DELPHES 3, A modular framework for fast simulation of a generic collider experiment,''
  JHEP {\bf 1402}, 057 (2014)
  \doi{10.1007/JHEP02(2014)057}
  [\arxiv{1307.6346} [hep-ex]].
  %%CITATION = doi:10.1007/JHEP02(2014)057;%%
  %598 citations counted in INSPIRE as of 10 Jan 2017
  
\bibitem{anti_kt}
  M.~Cacciari, G.~P.~Salam and G.~Soyez,
  %``The Anti-k(t) jet clustering algorithm,''
  JHEP {\bf 0804}, 063 (2008)
  \doi{10.1088/1126-6708/2008/04/063}
  [\arxiv{0802.1189} [hep-ph]].
  %%CITATION = doi:10.1088/1126-6708/2008/04/063;%%
  %4148 citations counted in INSPIRE as of 10 Jan 2017

\bibitem{fastjet}
 M.~Cacciari and G.~P.~Salam,
  %``Dispelling the $N^{3}$ myth for the $k_t$ jet-finder,''
  Phys.\ Lett.\  B {\bf 641}, 57 (2006)
  \doi{10.1016/j.physletb.2006.08.037}
  [\arxiv{hep-ph/0512210}];
  %%CITATION = PHLTA,B641,57;%%
 M.~Cacciari, G.~P.~Salam and G.~Soyez,
  %``FastJet User Manual,''
  Eur.\ Phys.\ J.\ C {\bf 72}, 1896 (2012)
  \doi{10.1140/epjc/s10052-012-1896-2}
  [\arxiv{1111.6097} [hep-ph]];
  %%CITATION = ARXIV:1111.6097;%%
  %\urlx{http://fastjet.fr}

\bibitem{recalibrate}
  C.~Guo, G.~Pleiss, Y.~Sun, and K.~Q.~Weinberger,
  %On Calibration of Modern Neural Networks 
  Proc. ICML (2017)
  [\arxiv{1706.04599} [cs.LG]].

\bibitem{platt_scaling}
  J.~Platt,
  %`Probabilistic outputs for support vector machines and comparisons to regularized likelihood methods. 
  Adv. Large Margin Classifiers {\bf 10(3)}, 61 (1999).

\bibitem{dropout}
  G.~E.~Hinton, N.~Srivastava, A.~Krizhevsky, I.~Sutskever, and R.~R.~Salakhutdinov,   %`Improving neural networks by preventing co-adaptation of feature detectors
  \arxiv{1207.0580} [cs.NE].

\bibitem{dropout_equiv}
  Y.~Gal and Z.~Ghahramani,
  %`Dropout as a Bayesian Approximation: Representing Model Uncertainty in Deep Learning 
  Proc. ICML (2016)
  [\arxiv{1506.02142} [stat.ML]].

\bibitem{top_review}
  T.~Plehn and M.~Spannowsky,
  %``Top Tagging,''
  J.\ Phys.\ G {\bf 39}, 083001 (2012)
  \doi{10.1088/0954-3899/39/8/083001}
  [\arxiv{1112.4441} [hep-ph]].
  %%CITATION = doi:10.1088/0954-3899/39/8/083001;%%
  %116 citations counted in INSPIRE as of 18 Mar 2019

\bibitem{jes}
  V.~Khachatryan {\it et al.} [CMS Collaboration],
  %``Jet energy scale and resolution in the CMS experiment in pp collisions at 8 TeV,''
  JINST {\bf 12}, no. 02, P02014 (2017)
  \doi{10.1088/1748-0221/12/02/P02014}
  [\arxiv{1607.03663} [hep-ex]];
  %%CITATION = doi:10.1088/1748-0221/12/02/P02014;%%
  %349 citations counted in INSPIRE as of 18 Apr 2019
  M.~Aaboud {\it et al.} [ATLAS Collaboration],
  %``Jet energy scale measurements and their systematic uncertainties in proton-proton collisions at $\sqrt{s} = 13$ TeV with the ATLAS detector,''
  Phys.\ Rev.\ D {\bf 96}, no. 7, 072002 (2017)
  \doi{10.1103/PhysRevD.96.072002}
  [\arxiv{1703.09665} [hep-ex]].
  %%CITATION = doi:10.1103/PhysRevD.96.072002;%%
  %241 citations counted in INSPIRE as of 18 Apr 2019

\bibitem{adv_ex}
  I.~J.~Goodfellow, J.~Shlens, and C.~Szegedy,
  %`Explaining and Harnessing Adversarial Examples
  \arxiv{1412.6572} [stat.ML].

\bibitem{puppi}
  D.~Bertolini, P.~Harris, M.~Low and N.~Tran,
  %``Pileup Per Particle Identification,''
  JHEP {\bf 1410}, 059 (2014)
  \doi{10.1007/JHEP10(2014)059}
  [\arxiv{1407.6013} [hep-ph]].
  %%CITATION = doi:10.1007/JHEP10(2014)059;%%
  %181 citations counted in INSPIRE as of 19 Mar 2019

\bibitem{softkiller}
  M.~Cacciari, G.~P.~Salam and G.~Soyez,
  %``SoftKiller, a particle-level pileup removal method,''
  Eur.\ Phys.\ J.\ C {\bf 75}, no. 2, 59 (2015)
  \doi{10.1140/epjc/s10052-015-3267-2}
  [\arxiv{1407.0408} [hep-ph]].
  %%CITATION = doi:10.1140/epjc/s10052-015-3267-2;%%
  %64 citations counted in INSPIRE as of 19 Mar 2019

\bibitem{MLpileup}
  P.~T.~Komiske, E.~M.~Metodiev, B.~Nachman and M.~D.~Schwartz,
  %``Pileup Mitigation with Machine Learning (PUMML),''
  JHEP {\bf 1712}, 051 (2017)
  \doi{10.1007/JHEP12(2017)051}
  [\arxiv{1707.08600} [hep-ph]].
  %%CITATION = doi:10.1007/JHEP12(2017)051;%%
  %36 citations counted in INSPIRE as of 19 Mar 2019
  J.~Arjona Martinez, O.~Cerri, M.~Pierini, M.~Spiropulu and J.~R.~Vlimant,
  %``Pileup mitigation at the Large Hadron Collider with Graph Neural Networks,''
  \arxiv{1810.07988} [hep-ph].
  %%CITATION = ARXIV:1810.07988;%%
  %5 citations counted in INSPIRE as of 19 Mar 2019


\end{thebibliography}
\end{document}